\documentclass[acmtog,screen,nonacm]{acmart}

\usepackage{colortbl}
\PassOptionsToPackage{table}{xcolor}
\usepackage{bm}
\usepackage{amsmath}
\usepackage{booktabs}
\usepackage[abs]{overpic}
\usepackage{nicefrac}
\usepackage{wrapfig}
\usepackage{soul}
\usepackage{colortbl}

\DeclareMathOperator*{\argmin}{arg\,min}

\newcommand{\refsec}[1]{Sec.~\ref{#1}}
\newcommand{\refeq}[1]{Eq.~\eqref{#1}}
\newcommand{\reffig}[1]{Fig.~\ref{#1}}

\newcommand{\reftab}[1]{Tab.~\ref{#1}}
\newcommand{\refapx}[1]{App.~\ref{#1}}

\newcommand{\system}{\textit{Dynamic Mode Decomposition}}

\newcommand{\koopman}{Koopman operator}

\AtBeginDocument{%
  }

\setcopyright{acmlicensed}
\acmJournal{TOG}
\acmYear{2025}
\acmVolume{44}
\acmNumber{4}
\acmArticle{}
\acmMonth{8}
\acmDOI{10.1145/3730826}

\begin{document}

\title{Fast Subspace Fluid Simulation with a Temporally-Aware Basis}

\author{Siyuan Chen}
\email{siyuangraphics.chen@mail.utoronto.ca}
\orcid{0009-0009-4309-862X}
\affiliation{
    \institution{University of Toronto}
    \city{Toronto}
    \country{Canada}
}
\affiliation{
    \institution{Shanghai Jiao Tong University}
    \city{Shanghai}
    \country{China}
}

\author{Yixin Chen}
\email{yixinc.chen@mail.utoronto.ca}
\orcid{0000-0001-7547-9587}
\affiliation{
    \institution{University of Toronto}
    \city{Toronto}
    \country{Canada}
}

\author{Jonathan Panuelos}
\email{jonathan.panuelos@mail.utoronto.ca}
\orcid{0009-0005-9643-0965}
\affiliation{
    \institution{University of Toronto}
    \city{Toronto}
    \country{Canada}
}

\author{Otman Benchekroun}
\email{otman.benchekroun@mail.utoronto.ca}
\orcid{0000-0001-6966-5287}
\affiliation{
    \institution{University of Toronto}
    \city{Toronto}
    \country{Canada}
}

\author{Yue Chang}
\email{changyue.chang@mail.utoronto.ca}
\orcid{0000-0002-2587-827X}
\affiliation{
    \institution{University of Toronto}
    \city{Toronto}
    \country{Canada}
}

\author{Eitan Grinspun}
\email{eitan@cs.toronto.edu}
\orcid{0000-0003-4460-7747}
\affiliation{
    \institution{University of Toronto}
    \city{Toronto}
    \country{Canada}
}

\author{Zhecheng Wang}
\email{zhecheng@cs.toronto.edu}
\orcid{0000-0003-4989-6971}
\affiliation{%
  \institution{University of Toronto}
  \city{Toronto}
  \country{Canada}
}

\renewcommand{\shortauthors}{Chen et al.}

\begin{abstract}
We present a novel reduced-order fluid simulation technique leveraging Dynamic Mode Decomposition (DMD) to achieve fast, memory-efficient, and user-controllable subspace simulation. We demonstrate that our approach combines the strengths of both spatial reduced order models (ROMs) as well as spectral decompositions. By optimizing for the operator that \emph{evolves} a system state from one timestep to the next, rather than the system state itself, we gain both the compressive power of spatial ROMs as well as the intuitive physical dynamics of spectral methods. The latter property is of particular interest in graphics applications, where user control of fluid phenomena is of high demand. We demonstrate this in various applications including spatial and temporal modulation tools and fluid upscaling with added turbulence. 

We adapt DMD for graphics applications by reducing computational overhead, incorporating user-defined force inputs, and optimizing memory usage with randomized SVD. The integration of OptDMD and DMD with Control (DMDc) facilitates noise-robust reconstruction and real-time user interaction. We demonstrate the technique's robustness across diverse simulation scenarios, including artistic editing, time-reversal, and super-resolution.

Through experimental validation on challenging scenarios, such as colliding vortex rings and boundary-interacting plumes, our method also exhibits superior performance and fidelity with significantly fewer basis functions compared to existing spatial ROMs. Leveraging the inherent linearity of the DMD formulation, we demonstrate a range of diverse applications. This work establishes another avenue for developing real-time, high-quality fluid simulations, enriching the space of fluid simulation techniques in interactive graphics and animation.

\end{abstract}

\ccsdesc[500]{Computing methodologies~Modeling and simulation}
\ccsdesc[500]{Computing methodologies~Physical simulation}

\keywords{Model-Reduction, Koopman Theory, Dynamic Mode Decomposition}


\begin{teaserfigure}
  \centering
  \includegraphics[width=1\linewidth]{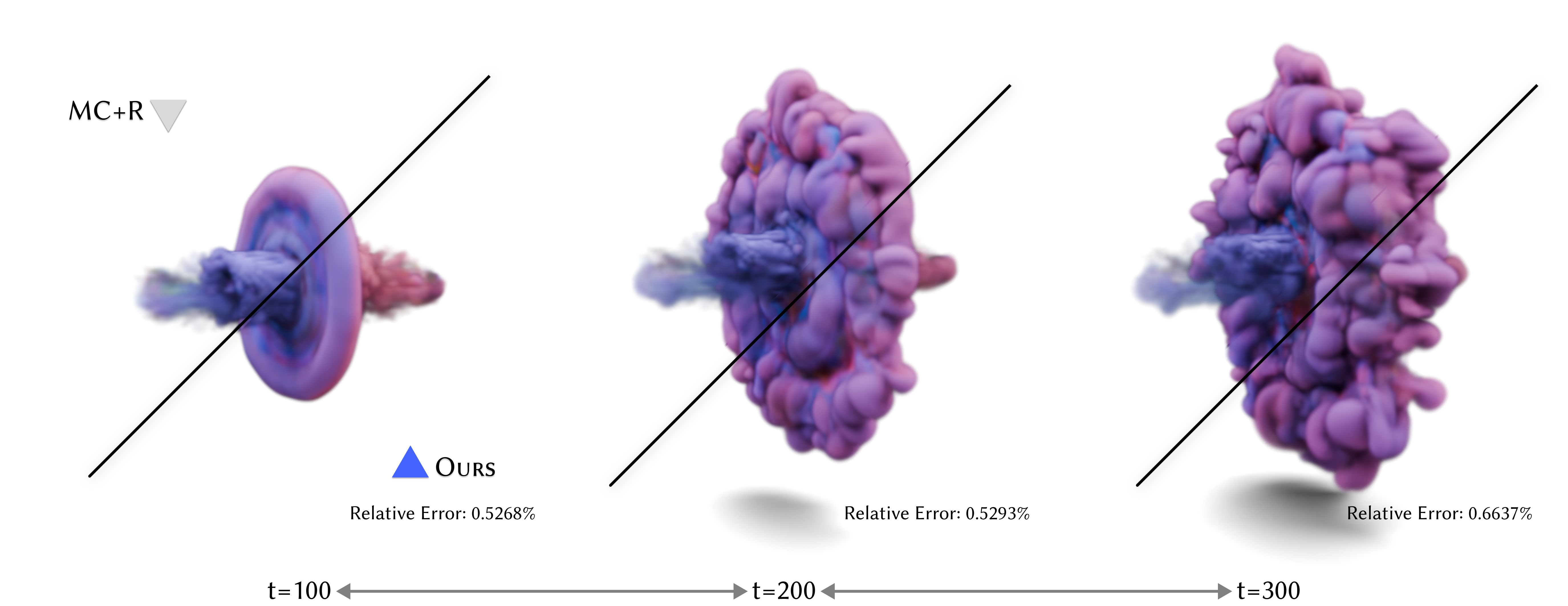}
  \caption{\textbf{Accurate Koopman-based Fluid Prediction of Vortex Ring Colliding.} We compare (\emph{above}) the ground truth MacCormack + Reflection (MC+R) solver \cite{zehnder2018advection} to (\emph{below}) our reduced Koopman-based reconstruction, which leverages a reduced-order temporally-aware subspace with $r=150$ basis functions. Our approach achieves efficient, near-perfect prediction of arbitrary future fluid states via the $k$-th power of the eigenvalue matrix. In this experiment, parameter $k=100$.} 
  \label{fig:teaser}
  \Description{}
\end{teaserfigure}

\maketitle

\begin{figure*}[!ht]
    \centering
    \includegraphics[width=\textwidth]{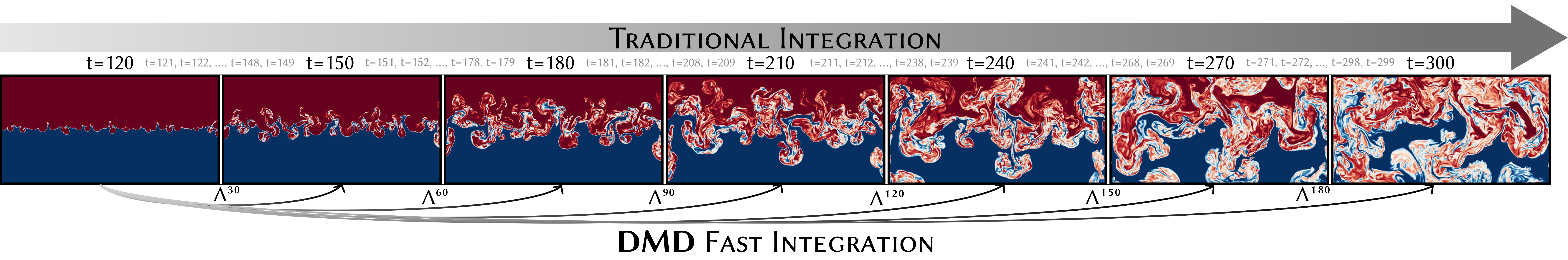}
    \caption{\textbf{Long-time Single Step Integration.} We demonstrate that our method can perform integration into arbitrary points in time via the \textit{exponential integration} of a single matrix (\refsec{sec:arbitrary_time_step}). Since the DMD operator is diagonal within the reduced basis, it is trivial to find the matrix that evolves the initial velocity field to the field at any point in time, significantly accelerating the integration as compared to traditional methods required by PCA. 
    Although the DMD operator is trained over the velocity field $\bm{u}$, we show the corresponding density field at each time point for clearer visualization.
    }
    \label{fig:dmdadvectioncomparison}
    \Description{}
\end{figure*}

\section{Introduction}
\label{sec:introduction}

Fast, responsive, and visually realistic fluid simulation remains a significant challenge in physics-based animation.
Traditional full-resolution simulations deliver highly detailed fluid flows but are computationally expensive, limiting their use to offline cinematic visual effects rather than interactive virtual reality applications.

Reduced-Order Models (ROMs) address this challenge by accelerating simulations through dimensionality reduction.
In the context of fluid dynamics, ROMs approximate the \emph{full, high-dimensional} fluid simulation solution space by effectively operating within a \emph{reduced, lower-dimensional} subspace.
The dynamics of the reduced fluid simulation are then obtained by projecting the high-dimensional solution space into the reduced space and solving the physical Partial Differential Equation (PDE) within this subspace.
As a result, the subspace needs to provide a compressed representation of the fluid state, while also capturing realistic and highly detailed fluid behavior.

Picking which subspace to use for simulation is non-trivial, and after committing to a subspace, using it for reduced simulation opens up separate difficulties. 
First, subspace fluid simulation is notorious for dissipating high-frequency detail, the kind that is commonly desired in turbulent flows. 
Second, it involves computing a static subspace that can generalize to a diversity of simulation states expected across a variety of scene interactions and configurations. 
Finally, even if one makes use of a linear subspace for simulation, the advection component of inviscid Euler equations is non-linear with respect to the fluid state, which requires full-space computation even with the use of a precomputed subspace.

Instead of committing to the static subspace methodology, we make use of Dynamic Mode Decomposition (DMD) \cite{schmid2010dynamic}, a modern flow analysis technique that takes an alternative simplification to the fluid simulation problem.
This formulation linearly approximates the \emph{\koopman{}},  which encodes the temporal evolution of the fluid flow, and performs reduction directly on this operator \cite{schmid2010dynamic}.
An immediate advantage of reduction on this operator is that the resulting subspace is imbued with information regarding the temporal dynamics of the flow.
This temporal awareness allows us to quickly evaluate the fluid flow at \emph{any} point in time directly, without performing any time integration or discrete fluid advection whatsoever, as shown in \reffig{fig:dmdadvectioncomparison}.

While DMD has shown great success for use in flow analysis from the engineering community, 
we are the first to show how it can be adapted for use in fluid animation applications in graphics, such as fluid editing, guiding, interaction, and artistic fluid control. 

Our key insight is to leverage the inherent spatio-temporal nature of the linear DMD operator. By encoding \emph{both} the \emph{spatial} nature of flows in the eigenvector basis and their \emph{temporal} evolution in the eigenvalues, each eigenvector-eigenvalue pair becomes representative of a distinct wave mode. 

This spectral-like decomposition allows for directable control of each mode separately, enabling new forms of creative expression, such as adjusting the amplitude of temporal frequency bands to control look-and-feel in interactive editing of fluid animation.

\paragraph{Contributions}
In summary, in this paper we:
\begin{itemize}
    \item introduce Dynamic Mode Decomposition (DMD) for fast control of fluid simulations with exceptional accuracy, computational speed, and memory efficiency;
    \item identify that the inherent spatiotemporal nature of the DMD operator enables control of each wave mode by modification of eigenvector-eigenvalue pairs;
    \item demonstrate DMD's versatility and practicality in the context of interaction and artistic control, such as frequency editing, time-reversal, super-resolution and simulation styling.
\end{itemize}
\section{Related Works}
\label{sec:related_works}

Following pioneering work by \citet{foster1996liquids}, \citet{stam1999stable}, and \citet{fedkiw2001visual}, the graphics community has made significant progress in the visual simulation of fluids.
Our method straddles literature from both fluid control and runtime acceleration. From foundational work, we aim to demonstrate that DMD arises naturally as a combination of spatial reduced-order modelling and spectral decompositions, a unique position that makes it well suited for fluid control applications.

\subsection{Fluid Control}
The chaotic nature of fluid behaviour makes artist-directed control a particularly difficult problem, as it often requires balancing accuracy, efficiency, and the user intuition. Of particular interest for us are applications in fluid snapshotting, where control \emph{frames} are provided that fluid flows into and out of. Numerous prior works have been done using optimization-based techniques, aiming to compute optimal control forces that guide fluid flows to match user-specified keyframes \cite{treuille2003keyframe, mcnamara2004fluid, thurey2009detail, inglis2017primal, pan2017efficient}. However, the high-dimensional optimization problems make most prior methods computationally expensive, limiting their application to real-time and interactive control. To further improve the efficiency, reduced-order fluid control methods have been proposed, such as frequency-aware force field reduction \cite{tang2021honey} and control with Laplacian Eigenfunctions \cite{chen2024fluid}. In this work, we demonstrate that the reversible nature of the DMD operator makes it particularly suitable for these problems. Time-reversibility has been prior explored in graphics \cite{oborn2018time}, but requires the solution of a modified Poisson problem. In comparison, we demonstrate that the DMD operator is trivially reversible. 

We also take influence from resolution-based smoke control \cite{nielsen2009guiding, sato2021stream}, which uses an artist-directed low-resolution input to produce a high-resolution simulation with secondary motion, such as smaller-scale turbulence. We demonstrate that the modes provided by DMD allow for direct application to these problems. We also similarly can preserve the input features via projection into the input velocity space similarly to \citet{nielsen2009guiding}, but show that our quadratic problem is precomputable and resolves into a simple matrix-vector multiply at runtime.

\subsection{Fast Fluid Simulation}
A major computational bottleneck in most Navier-Stokes (NS) solvers is the global pressure projection step, which is computationally expensive when solved directly, especially for high-resolution simulations. To mitigate this cost, iterative solvers such as Conjugate Gradient (CG)~\cite{saad2003iterative} and multigrid methods~\cite{briggs2000multigrid} are commonly employed to exploit the sparsity structure of the linear operators.

Avoiding this global pressure projection step is a key focus of real-time methods. To this end, Smoothed-Particle Hydrodynamics (SPH) \cite{muller2003particle}, and later Position-Based Fluids (PBF) \cite{macklin2013pbf}, have become the \emph{de facto} standard for GPU-based fluid simulation. 
More recently, the Lattice Boltzmann Method (LBM) has emerged as a powerful grid-based alternative \cite{chen1998lattice, chen2021gpu, li2020fast}. All these methods avoid a global solve, and are thus embarrasingly parallelizable and highly suitable for modern GPU architectures. 
In exchange, however, these methods often require extremely high resolutions and equivalently high memory consumption during runtime. By comparison, our method also avoids this global solve, evolving the system via a single matrix vector multiply, while simultaneously keeping the memory footprint low via reduction to a small basis.

\subsection{Spatial Order Reduction} \label{sec:related_roms}
In addition to the above parallelization approaches, reduced-order models present another promising direction for accelerating simulations. These methods aim to simplify the underlying computational models while preserving essential physical dynamics, thus improving speed and scalability. ROMs generally fall into two categories: \emph{data-free} and \emph{data-driven} methods.

\subsubsection{Data-free ROMs}
Data-free ROMs rely on simplified physics-based formulations to reduce computational complexity without requiring precomputed datasets. Laplacian Eigenfluids \cite{de2012fluid, liu2015model, cui2018scalable} has emerged as a class of data-free reduced-order methods for fluid simulations, representing states using a set of basis functions that best span a given spatial domain.

This approach demonstrates the feasibility of order reduction for fluid simulation. However, because their basis attempts to represent \emph{any} divergence-free field, handling intricate boundaries or capturing fine details often requires a large number of basis functions, and correspondingly large runtime and memory. If the general type of flow is known \emph{a priori}, which is often the case when an artist wants to apply simulation tools, these priors can be a powerful tool for improving computational efficiency.

\subsubsection{Data-driven ROMs}
Data-driven ROMs, in contrast, leverage statistical techniques and precomputed datasets to approximate complex fluid dynamics.
Essentially, precomputed data already encodes essential structures of the fluid flow; one only needs to extract that flow and simplify the representation elsewhere. 

\citet{treuille2006model} used principal component analysis (PCA) to find this reduced basis by minimizing reconstruction error given some target dimension. 
\citet{wicke2009modular} introduced spatial generalization by replacing the global basis with 'tiles' of local bases coupled with shared boundary bases.
A particular drawback for existing data-driven methods is the lack of intuition for these spatial bases. 
This makes it difficult to pick out individual bases and modify them to manipulate the resulting fluid flow, a key application that we aim to tackle.
General adoption has thus been limited for existing data-driven ROMs, and are largely relegated to playback. 

\subsection{Spectral Methods}
A key strength of spectral methods is the physical intuition of the basis functions. 
Whereas spatial order reductions construct basis functions based on the domain shape and are thus more physically intuitive for non-advecting phenomena such as elastic modes \cite{brandt2017compressed, sellan2023breaking}, spectral methods construct basis functions based on how the field evolves over time, 
leveraging the periodicity of fluid flows to represent the system's dynamics in the frequency domain.
Each mode thus corresponds to the propagation of a different wave, or coherent groups of waves. \citet{kim2008wavelet} leveraged this to add smaller-scale turbulence as a post-processing step, and \citet{chern2017inside} used this to directly modify vortex rings. 
This type of fluid control greatly motivates much of our application domains presented in \refsec{sec:application}.

Prominently among these are Fourier-based spectral methods. \citet{stam2001simple} solves stable fluids in Fourier space, though only in a periodic domains due to the global nature of Fourier bases. 
In contrast, wavelet-based methods localize these bases into individual waves \cite{jeschke2018water}, allowing for better expressivity as well as multiscale features. 
Fourier methods also greatly simplify advection, which becomes a linear transform in Fourier space \cite{chern2016schrodinger}, further speeding up computation. We adopt a similar scheme, where advection is encoded directly in our DMD operator.

\subsection{Dynamic Mode Decomposition}
\label{sec:dmd_related_works}
DMD is a data-driven method that approximates the Koopman operator, a linear operator that describes the temporal evolution of a dynamical system~\cite{schmid2010dynamic}.
Prior usage of DMD for fluids application has been limited, and have not been explored at all in computer graphics and creative applications.
Usage has primarily been limited to playback and short-term forecasting \cite{proctor2016dynamic}. 
Outside of fluid simulation, DMD has been used to process large datasets in high-dimensional systems \cite{williams2015data} and construct usable operators from noisy data \cite{askham2018variable}. This latter extension, called OptDMD, is of significant interest to us, as the turbulence inherent in fluid fields naturally produces noisy training data.

Similar to other data-driven ROMs, DMD constructs spatial bases that encode the different flows present in the training set. 
Because it approximates the \emph{operator} that evolves a state forward in time rather than the state itself, it also gains the advantages of spectral methods. 
In particular, each basis becomes associated with different turbulent scales, opening the door to fluid control and upscaling applications. By computing the eigenvalues and eigenvectors of the Koopman operator, DMD can identify the dominant modes of a system and predict its future behavior.

We propose the use of Dynamic Mode Decomposition (DMD) as a tool for compressing and manipulating fluid simulations, combining the spatial compressive power of data-driven methods with the physical intuition and control of spectral methods. 
\section{Koopman Operator and Dynamic Mode Decomposition}
\label{sec:background}

Consider a system of the form,
\begin{align}
    \label{eqn:autonomous_original}
    \frac{d\bm{u}}{dt} = \bm{f}(\bm{u}),
\end{align}
where $\bm{u}(\bm{x},t) \in \mathbb{R}^N$ is a time-varying state vector in $\bm{x}\in\mathbb{R}^{d},\ t\in\mathbb{R}$, and $\bm{f}:\mathbb{R}^N\to\mathbb{R}^N$ is a nonlinear operator. Since the right-hand side is not explicitly dependent on time, this \emph{autonomous} system evolves according to the current state $\bm{u}$, not when $\bm{u}$ is evaluated.

Suppose we collect a finite dictionary of observables, and stack them to yield the measurement vector,
\begin{align}
    \bm{g}(\bm{u}(t)) = \begin{bmatrix}
        g_1(\bm{u}(t)) \\ g_2(\bm{u}(t)) \\ \vdots \\ g_n(\bm{u}(t))
    \end{bmatrix} = \begin{bmatrix}
        \bm{u}(\bm{x}_1,t) \\ \bm{u}(\bm{x}_2,t) \\ \vdots \\ \bm{u}(\bm{x}_n,t)
    \end{bmatrix} \in \mathbb{R}^{Nn},
\end{align}
where $n$ is the total spatial degrees of freedom of our measurements, and we take $T\leq n+1$ independent measurements of $\bm{g}(\bm{u})$ at different times. There exists an operator $\bm{K}$ on this basis that applies the following \emph{linearized} time-stepping operation \cite{modern_koopman_theory},
\begin{align}
    \bm{g}_{k+1}=\bm{K}\bm{g}_k,\label{eq:linearization}
\end{align}
where the subscripts $\{k\in\mathbb{N}\ |\ 0\leq k < T\}$ indicate the time level of a given measurement. $\bm{K}$ is known as the restricted Koopman operator, and is guaranteed to exist and exactly recover the given input measurements as long as the number of time samples $T$ is fewer or equal to $n+1$, where $n$ is the size of the spatial degrees of freedom. Moving forward, for practical use, we only require the understanding that $\bm{K}$ exists, and we defer discussion of Koopman theory on why this is true to \refapx{sec:koopman_theory}.

\subsection{Approximating the Koopman Operator with Dynamic Mode 
Decomposition}
The restricted Koopman operator $\bm{K}$ has a basis with dimension on the number of degrees of freedom of the data (\emph{i.e.} the number of grid points). In most reasonable simulation resolutions, this is computationally intractable, considering $\bm{K}$ is almost certainly dense in this grid basis.

To address this, we follow a strategy similar to prior dimensionality reduction approaches such as PCA \cite{kim2013subspace,treuille2006model}, and instead approximate the Koopman operator using \system{} (DMD) \cite{schmid2010dynamic}, constructing a \textit{reduced, low-dimensional} basis to represent the function space spanned by the input observables. Note that while $\bm{K}$ will exactly reproduce input measurements (and reproduce the continuous system in \refeq{eqn:autonomous_original} in a least-squares sense on the finite measurement basis), the dimensionality reduction offered by DMD serves to approximate $\bm{K}$, and thus will introduce extra error in exchange for computational tractability.

As illustrated on the right, a key distinction between PCA and DMD is their relationship to time.  Here, two consecutive snapshots are shown as two black dots (states recorded by frames) connected by a timeline.  Conventional \textit{PCA} is data order insensitive as its singular-value decomposition depends only on the covariance of the snapshot set, so re-ordering the frames leaves the low-rank surface where the projected states (red dots) reside unchanged.
\begin{wrapfigure}{r}{0.7\columnwidth}
    \hspace{-8pt}
    \includegraphics[width=0.7\columnwidth]{figure/pca_vs_dmd.png}
    \vspace{-8pt}
    \label{fig:pca_vs_dmd}
\end{wrapfigure}
\textit{DMD}, in contrast, fits an operator~$\bm{K}$ that advances the earlier state to the next predicted state (hollow red dot), as indicated by the red arrow. Because this least-squares fit relies on aligned pairs of samples, shuffling the data alters~$\bm{K}$.  This built-in temporal sensitivity lets DMD capture system dynamics rather than merely the geometry of the solution manifold, making it better suited for modelling time-evolving systems.
Intuitively, PCA models a subspace of the solution space, whereas DMD models the operator that maps a state to the next—an objective that naturally requires temporal information.

Given two \textit{time-shifted} sequence of snapshots $\bm{X}$ and $\bm{X'}$, a discrete Koopman operator is defined by:
\begin{equation}
    \label{eqn:reduced_koopman}
    \begin{aligned}
        &\argmin_{{\bm{K}}} \; \|\bm{X}^\prime - {{\bm{K}}} \bm{X}\|_F,  \\
        \text{where } \; 
        \bm{X} &= \underbrace{
            \begin{bmatrix}
                \bm{u}(0) & \bm{u}(1) & \cdots & \bm{u}(T-1) \\
            \end{bmatrix}}_{T \;\text{Frames}
        },
        \\
        \bm{X}^\prime &=
        \underbrace{
            \begin{bmatrix}
                \bm{u}(1) & \bm{u}(2) & \cdots & \bm{u}(T) \\
            \end{bmatrix}}_{T \;\text{Frames}
        },
    \end{aligned}
\end{equation}
Intuitively, this optimizes $\bm{K} \in \mathbb{R}^{(Nn) \times (Nn)}$ to minimize the difference between the next states $\bm{u}(t+\Delta{}t)$ and its prediction of the next states ${\bm{K}}\bm{u}(t)$ under a linear model.

We can solve this minimization as follows:
\begin{equation}
    \label{eqn:reduced_koopman_solution}
    {\bm{K}} = \bm{X}^\prime \bm V \bm \Sigma^{-1} \bm U^T,
\end{equation}
where $\bm{X} = \bm{U} \bm{\Sigma} \bm{V}^T$ is the singular value decomposition (SVD) of the snapshot matrix $\bm{X}$, with $\bm{V} \in \mathbb{R}^{(Nn) \times r}$ being the left eigenvectors, $\bm{\Sigma}$ being the eigenvalues matrix, and $\bm{U} \in \mathbb{R}^{(Nn) \times r}$ being the right eigenvectors.
We can truncate this eigensystem, taking the top $r$ singular values corresponding vectors to form the truncated \koopman{} $\tilde{\bm{K}}$. 
We can then project this truncated operator to the reduced basis $\bm{U}^T\tilde{\bm{K}}\bm{U} =  \bm{U}^T\bm{X}^\prime \bm V \bm \Sigma^{-1} \bm U^T\bm{U} = \hat{\bm{K}}$ to produce the reduced Koopman operator $\hat{\bm{K}}$.

We can see that as long as a mode is representable in the reduced space, $\hat{\bm{K}}$ and $\tilde{\bm{K}}$ share that mode's eigenvalue, $\hat{\bm{K}}\bm w_i = \lambda_i \bm w_i$, with the eigenvector simply being $\tilde{\bm{K}}$'s eigenvector projected onto the reduced space, $\bm w_i = \bm U^T \bm \phi_i$.
Here, $\lambda_i$ is a mode's eigenvalue, $\bm \phi_i$ is its corresponding eigenvector in full space, and $\bf w_i$ is its projection onto reduced space. 
We thus can find a spectral decomposition of the truncated Koopman operator: $\tilde{\bm{K}} = \bm{\Phi}\bm{\Lambda}\bm{\Phi}^*$.

Since applying the truncated \koopman{} $\tilde{\bm{K}}$ on the full space has complexity $\mathcal{O}(N^2n^2)$, we \textit{project} the full-space velocity field $\bm{u}$ onto the reduced space spanned by the basis $\bm{\Phi}$:
\begin{equation}
    \label{eqn:projection}
    \bm{u}(t + \Delta t) = \tilde{\bm{K}} \bm{\Phi} \bm{\Phi}^+ \bm{u}(t) = \tilde{\bm{K}} \bm{\Phi} \bm{z}(t),
\end{equation}
where $\bm{\Phi}^+ = (\bm{\Phi}^* \bm{\Phi})^{-1}\bm{\Phi}^* \in \mathbb{R}^{(Nn) \times r}$ is the Moore-Penrose pseudoinverse of $\bm{\Phi}$ and $\bm{z}(t) = \bm{\Phi}^+ \bm{u}(t) \in \mathbb{R}^r$ is the reduced state of the fluid system at time $t$. This projection step reduces the complexity to $\mathcal{O}(Nnr)$.

Notice now that taking $\bm{\Lambda} = \bm{\Phi}^*\tilde{\bm{K}}\bm{\Phi}\in \mathbb{R}^{r\times r}$ gives exactly a matrix that advances the reduced state forward in time by $\Delta{}t$:
\begin{equation}
    \label{eqn:reduced_koopman_simulation}
    \bm{z}(t + \Delta t) = \bm{\Lambda} \bm{z}(t),
\end{equation}
We note that $\bm{\Lambda}$ is the diagonal matrix of eigenvalues of $\hat{\bm{K}}$, and $\bm{\Phi}$ are the corresponding eigenvectors. That is to say, $\tilde{\bm{K}}=\bm{\Phi}\bm{\Lambda}\bm{\Phi}^*$ is exactly the spectral decomposition of the reduced Koopman operator. This low-rank model allows linear simulation of the reduced state, offering an efficient surrogate for the nonlinear underlying system.

From here, we can once again simply apply the basis $\bm{\Phi}$ to return back to full space:
\begin{equation}
    \label{eqn:reduced_koopman_projection}
    \bm{u}(t + \Delta t) = \bm{\Phi} \bm{\Lambda} \bm{z}(t).
\end{equation}

We would like to point out that despite both $\bm{\Lambda}$ and $\tilde{\bm{K}}$ are are referred to as Koopman operators in the literature, we remark that in our formulation they act on a subspace of \textit{linear} observables derived from state components. Therefore, the dynamics we capture are strictly linear, and our model is best interpreted as a linear approximation of the full underlying nonlinear system.

$\bm{\Lambda}$ in particular gives particular insight into the theory; notice here that it acts on the function space of $g=\bm{\Phi}$. Each eigenfunction is thus a special observable, which clearly behaves linearly by definition. Since we know that $\bm{\Phi}^*$ maps observables to observables, $\tilde{\bm{K}}$ must necessarily also be a Koopman operator, this time acting on the identity operator (or more accurately, the $\bm{u}$ observable in Hilbert space).

In addition, the $\bf \Lambda$ representation exposes the modes in an easily manipulatable manner. As a Koopman operator, it represents the time evolution of some observable. By being a diagonal matrix of eigenvalues, it turns out that these observables are exactly spatial modes that rotate with a particular frequency given by the imaginary part of the eigenvalues.
\begin{figure*}[ht]
    \centering
    \includegraphics[width=\linewidth]{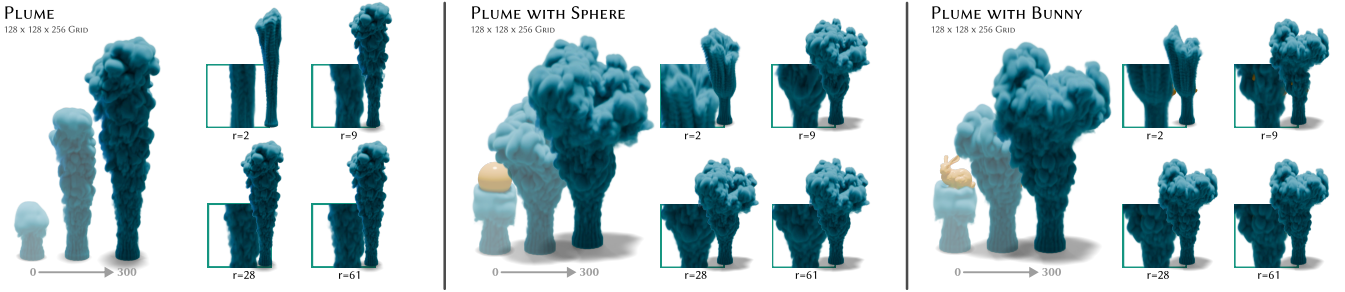}
    \caption{\textbf{Reconstruction of 3D Plume Simulations.} From left to right: standalone plumes, plumes interacting with a sphere, and plumes interacting with a bunny. For each configuration, the last frame of the original MacCormack \cite{selle2008unconditionally} fluid simulation is shown on the left, alongside the last frame of our subspace simulation with SVD rank ranging from $r = 2$ to $r = 61$. Remarkably, the fluid dynamics demonstrate strong resilience to low-rank bases, highlighting a key advantage of our proposed reduced-order pipeline. Additionally, these scenarios illustrate the robustness of our method in handling increasingly complex boundary conditions. The reconstruction quality improves as the number of basis functions increases, enabling more accurate capture of finer details around the boundaries. While reasonable results are achieved with a small basis (r=9), increasing the basis significantly enhances the fidelity of the simulations. More details on the temporal evolution of the flow and other basis configurations can be found in the \reffig{fig:appendix_plume}, \reffig{fig:appendix_sphere} and \reffig{fig:appendix_bunny}.} 
    \label{fig:basis_evaluation}
    \Description{}
\end{figure*}

\section{Adapting DMD for Graphics Applications}
\label{sec:methodology}

\subsection{Constructing Training Data}
We point out that the system solved by \refeq{eqn:reduced_koopman} is agnostic to the structure of the input state $\bm{u}$. As stated, the DMD operator learns a mapping that evolves the state $\bm{u}$ from one point in time to its state at a future moment, and the natural description of a fluid state is to represent $\bm{u}$ as a stacked vector of fluid velocities, which implies that the DMD operator learns the \emph{PDE} that governs the evolution of the fluid velocities over time.

For NS solvers, the evolution of a fluid is described by the inviscid Euler equations,
\begin{equation}
    \label{eqn:euler_equations}
    \begin{cases}
        \frac{\partial \bm{u}}{\partial t} + (\bm{u} \cdot \nabla) \bm{u} = -\frac{1}{\rho} \nabla p + \bm{g} \\
        \nabla \cdot \bm{u} = 0
    \end{cases}
\end{equation}
where $\bm{u}$ is the velocity field, $p$ is the pressure, $\rho$ is the density, and $\bm{g}$ is the gravitational force. We discretize the domain $\Omega$ with the staggered grid, a standard approach in computer graphics \cite{harlow1965numerical}, where the velocity field $\bm u$ is defined at the cell faces and the pressure $p$ is defined at the cell centers. This thus provides snapshots of the fluid velocity field $\bm{u}(t)$, at any point in time $t$. Note that $\bm{u}$ is a vector of size $\mathbb{R}^{Nn}$. This vector space henceforth will be our \emph{fullspace}, as it includes spatially every point in our full resolution simulation.

Since our training data can be generated from any simulation method (or even observed from the real world), we utilize several fluid simulation algorithms to construct our dataset. Specifically, we include the standard stable fluids \cite{stam1999stable}, the MacCormack \cite{selle2008unconditionally} (MC), the MacCormack + Reflection \cite{zehnder2018advection} (MC+R) and the Lattice Boltzmann Method with the Bhatnagar-Gross-Krook collision model \cite{chen1998lattice} (LBM-BGK).

\subsection{Learning from Noisy Data with Nonlinear Optimization}

Fluid behaviour is inherently chaotic, particularly in highly turbulent systems. Additionally, discretization of such states leads to noisy data, particularly as structures approach the Nyquist limit of the sampling grid. Further, simulation speed and artist directability is of primary concern in graphics application above simulation accuracy, often leading to simulations with CFD-condition-violating large timesteps and early termination of iterative algorithms. This further leads to degraded simulation results. As such, we note that input fluid data in general will be highly noisy.

Standard DMD, as shown in \refeq{eqn:reduced_koopman_solution}, directly fits the data from the $i$-th frame to the ($i+1$)-th frame without considering the (potentially coherent) relationships between the $i$-th frame and the ($i+2$)-th or subsequent frames. This short time horizon makes it highly sensitive to noise, both spatially as the fluid state changes rapidly at the Nyquist limit, and at the operator level. As a result, standard DMD imposes high requirements on data quality. 

To address this issue, we choose to use OptDMD \cite{askham2018variable} for our graphics applications. 
OptDMD processes all snapshots simultaneously, significantly reducing the impact of noise by looking at the signal over a longer time horizon. 
By considering the $i$-th frame, the ($i+1$)-th frame, the ($i+2$)-th frame, and so on together, the influence of random noise is mitigated. 
To achieve this, OptDMD transforms the training of the reduced Koopman operator into an exponential data fitting problem and utilizes the variable projection method to solve this optimization problem.
Similar to the procedure described in \refsec{sec:background}, we first perform Singular Value Decomposition (SVD) on the snapshot matrix $\bm{X}$ (see \refeq{eqn:reduced_koopman}) to obtain $\bm{U}, \bm{\Sigma}, \bm{V}$ such that $\bm{X} = \bm{U} \bm{\Sigma} \bm{V^T}$. Here, $\bm{\Sigma}$ is the diagonal singular value matrix, and $\bm{U}, \bm{V}$ are orthonormal matrices whose columns are the left and right singular vectors, respectively.

With this decomposition, we can construct the optimization problem as:
\begin{equation}
    \label{eqn:variable_projection}
    \argmin_{\bm{\alpha}, \bm{B}} \; \|\bm{\bar{V}}\bm{\Sigma} - \bm{\Phi}(\bm{\alpha})\bm{B}\|_F
\end{equation}
where $\bm{\bar{V}}$ denotes the element-wise complex conjugate of $\bm{V}$, $\bm{\Phi}(\bm{\alpha})$ is the parameterized basis matrix with optimizable parameters $\bm{\alpha}$, and $\bm{B}$ is the coefficient matrix corresponding to the basis matrix.
Here $\bm{\Phi}(\bm{\alpha})\in\mathbb{C}^{T\times r}$ is the Vandermonde matrix whose
$(i,j)$-entry is $[\bm{\Phi}(\bm{\alpha})]_{ij}=\alpha_j^{\,i-1}$ for $i=1,\dots,T$ and $j=1,\dots,r$; the vector $\bm{\alpha}=[\alpha_1,\dots,\alpha_r]^\top$ collects the (discrete-time) DMD eigenvalues $\bm\alpha = \bm\Lambda$, initialized from the eigenvalue matrix $\bm\Lambda$ of a standard DMD solve (\refeq{eqn:reduced_koopman}).

Once the coefficient matrix $\bm{B}$ is estimated, we could compute each column of the basis matrix $\bm{\Phi}$ using:
\begin{equation}
     \label{eqn:phi_optdmd}
    \bm{\phi_i} = \frac{\bm{U_r}\bm{B^T}(:, i)}{\|\bm{U_r}\bm{B^T}(:,i)\|_2}
\end{equation}
where $\bm{\phi_i}$ denotes the $i$-th column of matrix $\bm{\Phi}$.

\subsection{Memory Overhead Optimization}
DMD can effectively reconstruct datasets, but its training process requires a significant amount of memory due to the SVD. When the dataset is large, it becomes necessary to store an $Nn \times Nn$ matrix in memory, where $Nn$ represents the dimensionality of a snapshot, thus demonstrating a significant memory requirement. 

We note that previous DMD literature either has much smaller datasets afforded by 2D data, or perform training on large clusters \cite{schmid2010dynamic, proctor2016dynamic, askham2018variable, sashidhar2022bagging}. Uniquely in graphics, we expect algorithms to be able to simulate 3D examples (i.e. large degrees of freedom) but run on consumer hardware (i.e. limited memory). 

To reduce the memory requirements of DMD, we employ randomized SVD. Randomized SVD approximates the range of the original matrix $\bm{X}$ by constructing a matrix $\bm{Q}$. This is done by multiplying a randomly initialized low-dimensional matrix $\bm{Q}$ with the original matrix $\bm{X}$ and performing decompositions iteratively until a stable vector matrix $\bm{Q}$ is obtained. The goal is to ensure that  $\bm{X} \approx \bm{Q}\bm{Q^T} \bm{X}$. We then construct the matrix  $\bm{B} = \bm{Q^T} \bm{X}$, which applies to the space spanned by $\bm{Q}$. Traditional SVD can thus be performed on the much smaller $\bm{B}$ to obtain  $\bm{B} = \bm{U} \bm{\Sigma} \bm{V^T}$, and finally be used to approximate $\bm{X} \approx \bm{Q} \bm{U} \bm{\Sigma} \bm{V^T}$.

\begin{figure}[!ht]
    \centering
    \includegraphics[width=1\columnwidth]{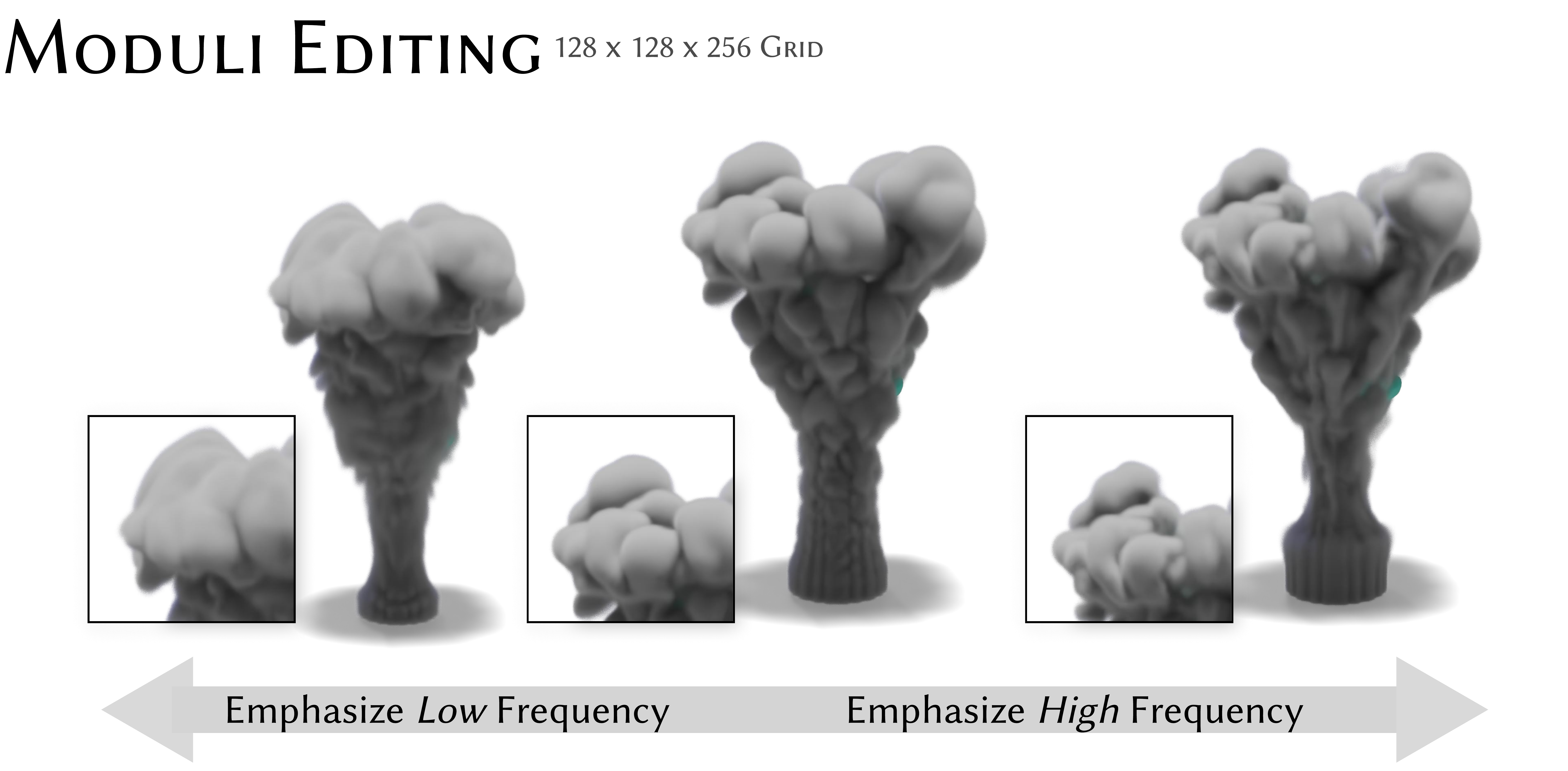}
    \caption{\textbf{Editing Temporal Dynamics of the Plume with Bunny with the Koopman Operator Approximation}. In this experiment, we demonstrate the impact of changing the moduli of low-frequency and high-frequency modes in a 4:1 (left) and 1:4 (right) ratio. As mentioned in \refsec{sec:editing}, in real applications, users can modify and manipulate the dynamic of different scales of vorticity by adjusting the reduced-order parameters.}
    \label{fig:bunny_editing}
    \Description{}
\end{figure}

\subsection{Arbitrary Time Step}
\label{sec:arbitrary_time_step}
Since the reduced \koopman{} is a linear operator, as shown in \reffig{fig:dmdadvectioncomparison} we could easily pre-compute the power of the reduced \koopman{} and apply it to a reduced state $z$, enabling \textit{large} time step $k$ times of the time step $\Delta t$ that $\bm A$ is trained on:
\begin{equation}
    \label{eqn:large_time_step}
    \bm{u}(t + k\Delta t) = \bm{\Phi} \bm{\Lambda}^k z(t)
\end{equation}
where $k$ is an integer of how many time steps we want to simulate forward. This can map any state at time $t$ to the state at time $t + k\Delta t$ in \textbf{one} matrix multiplication.

Notice additionally that because the operator maps velocity fields to velocity fields, advection of the velocity variable is achieved for \emph{free}. Nonlinear advection of the field is encoded directly in the operator.

\subsection{Boundary Conditions and Divergence Free Constraint}
The truncated Koopman operator $\tilde{\bm{K}}$ is trained on a full-space velocity field $\bm{u}$ which satisfies some boundary conditions and the divergence free constraint. We follow a similar proof to that provided by \citet{treuille2006model} to demonstrate that the Koopman operator also satisfies the space boundary conditions and constraints. 

First, the boundary conditions and constraints can be represented as some linear constraint matrix $\bm C$ on the time series data $\bm X$ such that $\bm{C} \bm{X} = 0$.
Any column $b_i$ in $\bm{U_r}$ by definition satisfies $\bm{X}\bm{X^T}\bm{b_i} = \lambda_i \bm{b_i}$. We can then conclude that $\bm{C}\bm{X}\bm{X^T}\bm{b_i} = \lambda_i\bm{C}\bm{b_i} = 0$, and thus claim that $\bm{b_i}$ satisfies the constraint matrix as long as $\lambda_i \ne 0$.

As this is true for all columns, then $\bm{C}\bm{U_r} = 0$ and consequently $\bm{C}\bm{\Phi} = 0$. This means that any velocity field $\bm u$ reconstructed from its reduced representation satisfies any linear constraint satisfied by the initial data.

\subsection{Encompassing External Forces in Reduced Space}

The OptDMD method in CFD can only reconstruct the dataset and cannot respond to user inputs of new external forces. However, in graphics applications, it is very common to edit or manipulate existing datasets, such as adding or reducing forces. To address this, we combine DMD with Control (DMDc) \cite{proctor2016dynamic} and OptDMD, marking the first instance of integrating these two methods. This integration allows user-input external forces to be incorporated into the OptDMD framework. Our improved framework is as follows:
\begin{equation}
    \label{eqn:external_forces}
    \bm{z}(t + \Delta t) = \bm{\Lambda} \bm{z}(t) + \bm{\Phi^+} \sum_{i = 1}^m\bm{B_i} \bm{q_i(t)} \Delta t + \bm{\Phi^+} \sum_{j = 1}^n\bm{f_j(t)} \Delta t
\end{equation}
where $\bm B_i$ and $\bm q_i$ can be constructed differently based on the settings of various scenarios. $\bm f_j$ $\in \mathbb{R}^{(Nn)\times 1}$ represents a user-modifiable external force. For example, in the 2D standalone plume scenario, we can define $\bm q_1 \in \mathbb{R}^{M \times 1}$ as the density field, where $M$ represents the number of grid cells, and rewrite its impact on the velocity field in matrix form $\bm B_1 \in \mathbb{R}^{(Nn) \times M}$. Similarly, we can define $\bm q_2 \in \mathbb{R}^{M \times 1}$ as the temperature field, and rewrite its impact on the velocity field in matrix form $\bm B_2 \in \mathbb{R}^{(Nn) \times M}$. These effects can then be projected onto the reduced space as $(\bm \Phi^+ (\bm B_1 \bm q_1 + \bm B_2 \bm q_2))$.
\begin{figure}[!ht]
    \centering
    \includegraphics[width=0.8\columnwidth]{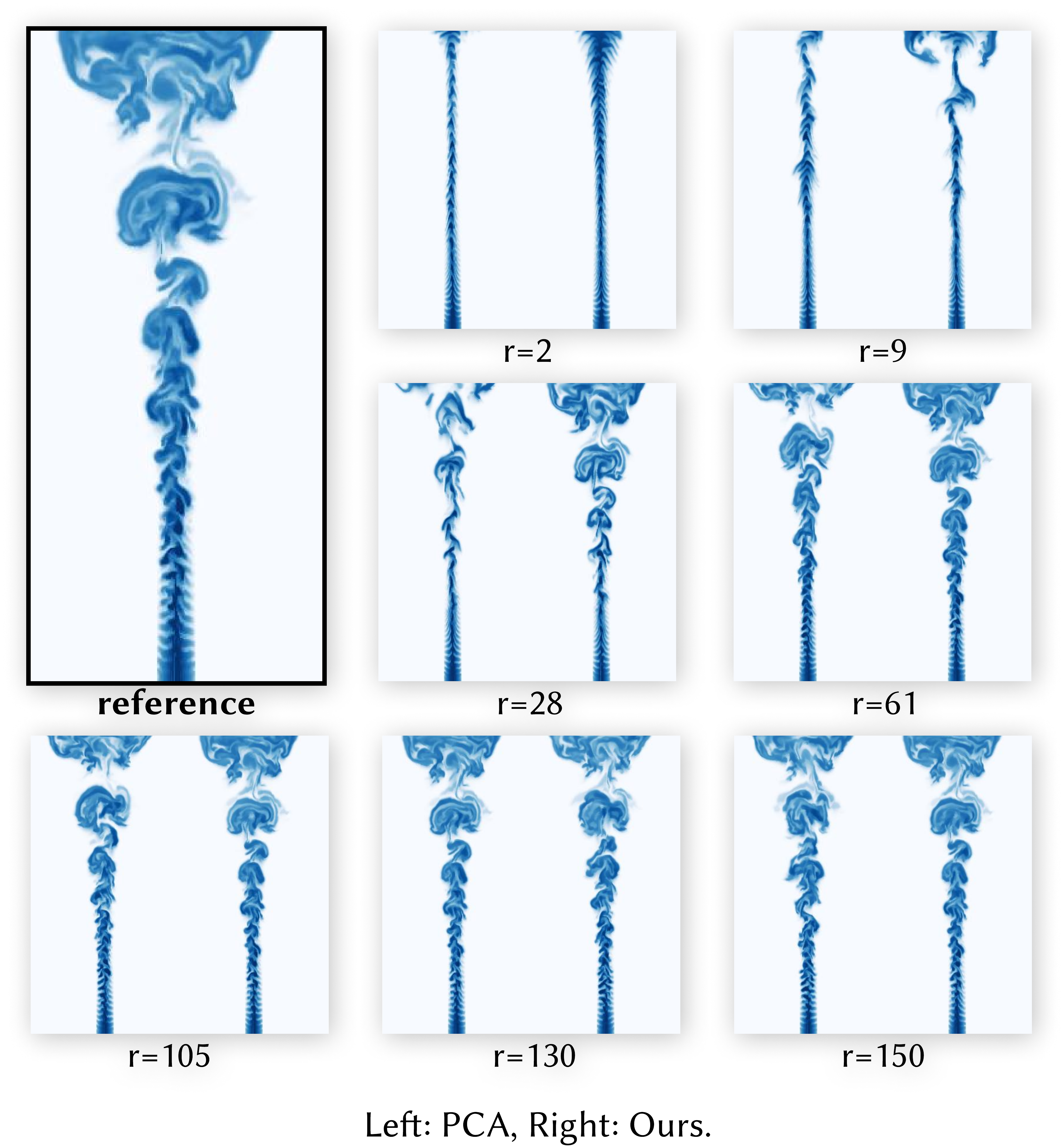}
    \caption{\textbf{Comparison of PCA and Our Method for Low-Rank Flow Reconstruction.} The leftmost column shows the reference high-resolution simulation, while the right grid presents reconstructions at different rank truncations ($r$). Each pair in the grid compares Principal Component Analysis (PCA) (left) and our method (right). Lower ranks ($r=2,9$) fail to capture large-scale turbulence while increasing \(r\) improves accuracy. Our method retains more detailed structures at lower ranks than PCA, demonstrating improved efficiency in capturing complex flow dynamics.}
    \label{fig:reconstruction}
    \Description{}
\end{figure}

\section{Evaluation and Results}
\label{sec:results}
In this section, we demonstrate the results of our approach on baseline examples and provide detailed numerical evaluations.
We conduct all numerical experiments on a Linux workstation equipped with a 2.10GHz 32-core Intel CPU featuring 120GB of RAM. We implement our method exclusively in Python, and our DMD implementation is a heavily optimized, memory-efficient version of \texttt{PyDMD} \cite{ichinaga2024pydmd}. We use \texttt{Taichi} \cite{hu2019taichi} for interactive examples and parallelization of DMD model inference, and \texttt{NumPy} \cite{harris2020array} for numerical validation. Please refer to \reftab{tab:experiment_setup} for the experiment setup of the examples presented in this section.

\subsection{Comparison of 2D Plume Reconstruction}
To evaluate the reconstruction performance of our method in comparison with the prior work \cite{kim2013subspace}, we conducted experiments on a 2D plume scenario using varying numbers of basis functions: $2, 9, 28, 61, 105, 130$, and $150$; shown on \reffig{fig:reconstruction}. With only 9 basis functions, our method is already capable of reconstructing the contour of the plume's top, while the prior method can only capture basic upward motion of the plume structure without any vortical structure details. This shows the superior reconstruction capability of our method even at low ranks. When using $28$ basis functions, our method closely matches the ground truth, whereas the prior method still misses many details, particularly in the upper regions of the plume. When the number of basis functions is increased to 61, our method achieves a near-perfect reconstruction of the ground truth, and the prior method requires 130 basis functions to produce a similar level of detail. These results effectively demonstrate that our method can achieve better reconstruction with significantly fewer basis functions.

\begin{table}[ht]
    \centering
    \rowcolors{2}{white}{gray!20}
    \begin{tabular}{l|c}
        \multicolumn{2}{l}{\textbf{\Large{\textsc{Runtime: 2D Plume}}} \;\textsc{\textcolor{gray}{256 $\times$ 512 Grid}}\; ($\triangleright$ \reffig{fig:reconstruction})} \\
        \hline
        \textbf{Fullspace Solve} & 754 ms  \\
        \hline
        \textbf{Subspace Solve} & \\
        \hspace{1em} PCA-based \shortcite{kim2013subspace} & 16.15 ms (47$\times$) / 0.95 ms (794$\times$) \textsuperscript{\textdagger}   \\
        \hspace{1em} \textbf{Ours} & \textbf{8.9 $\mu$s (84,719$\times$)}   \\
        \hline
        \textbf{Precomputation} &  \\
        \hspace{1em} PCA-based \shortcite{kim2013subspace} & 114 s  \\
        \hspace{1em} \textbf{Ours} & \textbf{79 s} \\
    \end{tabular}
    \caption{\textbf{Breakdown of Experiment Runtime.} 
    The table compares the runtime and precomputation costs of the full-space ground truth simulation, the PCA-based approach \shortcite{kim2013subspace} and our method on the 2D plume example (\reffig{fig:reconstruction}). Leveraging the linear \koopman{}, our method achieves significantly faster runtime by requiring only a single matrix multiplication for each reduced-space simulation step. 
    \\\textsuperscript{\textdagger} = with/without external force.}
    \label{tab:dmd_runtime}
\end{table}

\subsection{Generalization With Vorticity Confinement}
\label{sec:generalization}
\begin{figure}[!ht]
    \centering
    \includegraphics[width=1\columnwidth]{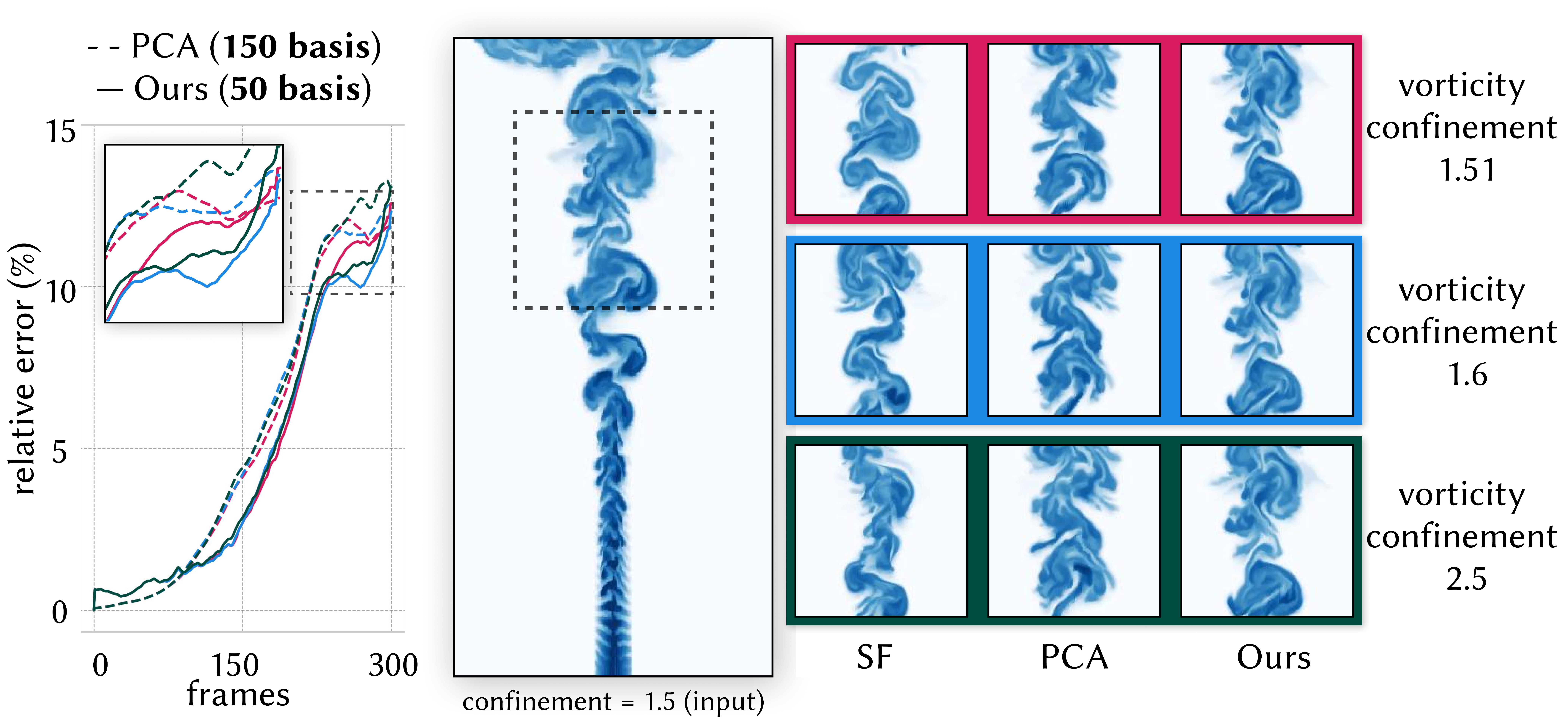}
    \caption{\textbf{Comparison of Relative Error and Vorticity Confinement Between PCA \shortcite{kim2013subspace} and DMD (Ours).} Left to right: (1) relative error over time for \textit{PCA (150 basis)} (dashed) and \textit{Ours (50 basis)} (solid), showing comparable or lower error with fewer basis functions. (2) reference high-resolution simulation with vorticity confinement $1.5$, with a zoomed-in region marked. (3) comparison of the zoomed region in (2) under novel unseen vorticity confinement force $1.51$ (magenta), $1.6$ (blue), $2.5$ (dark green).}
    \label{fig:generalization}
    \Description{}
\end{figure}

To further evaluate generalization and artist-directability of our method, we revisited the 2D plume example following the experimental setup from the prior work \cite{kim2013subspace}. Specifically, we incorporated vorticity confinement \cite{fedkiw2001visual} into the original MacCormack solver~\cite{selle2008unconditionally} and then applied our method to the modified solver. For training, the vorticity confinement value was set to 1.5, and we tested both our approach and the prior method at vorticity confinement values of 1.51, 1.6, and 2.5. To quantify generalization efficacy, we calculated the relative error of the velocity field with respect to ground truth MacCormack\cite{selle2008unconditionally} simulation results. The results, as shown in \reffig{fig:generalization}, reveal that our method achieves a lower relative error compared to the prior method. Notably, the prior method required 150 basis functions, whereas our method achieved comparable results with only 50 basis functions. These findings highlight the reliability and effectiveness of our Koopman-based approach in adapting to changes in simulation parameters.

\subsection{3D Plume Baselines}
To benchmark our Koopman-based fluid simulation pipeline in 3D scenarios, we present simulation results across three progressively complex scenarios: a standalone plume, a plume interacting with a sphere, and a plume interacting with a bunny. All original simulations were generated using the MacCormack solver \cite{selle2008unconditionally} at a resolution of $128 \times 128 \times 256$. We demonstrate the reconstruction results using varying numbers of basis functions. These cases were designed to benchmark the method’s ability to handle increasingly intricate fluid dynamics, from simple turbulence in the standalone plume to complex boundary interactions with the sphere and bunny.

\paragraph{\textbf{Plume}}
The standalone plume serves as a fundamental benchmark, as tested in prior work \cite{kim2013subspace}. \reffig{fig:basis_evaluation} highlights the effectiveness of our Koopman-based method in simulating a plume without solid obstacles. While \cite{kim2013subspace} experienced significant high-frequency dissipation at reduced ranks, our method well captures the swirling and rising behavior, closely matching the original MacCormack simulations even with a small number of basis functions ($r=9$). Increasing the basis amount to $r=28$ or $r=61$ further enhances the reconstruction, better preserving turbulent vortices and fine flow structures. These results demonstrate the robustness of our method in capturing the essential dynamics of the fluid simulation, maintaining high fidelity even at low ranks.

\paragraph{\textbf{Plume with Sphere and Bunny}}
Building on the baseline, we introduce two more scenarios to evaluate the robustness of our method in handling complex boundary conditions. When using a small set of basis ($r=9$), our Koopman-based method still achieves a reasonable reconstruction of the overall flow dynamics. However, finer details, especially those near boundaries, are less accurately represented. Notably, at $r=9$, there exist noticeable discrepancies in regions around the bunny, which will be resolved as the number of basis functions increases. This enhancement demonstrates how additional basis functions help the Koopman operator to reconstruct higher-frequency components of the flow and better capture complex boundary interactions.

\subsection{3D Colliding Vortex Rings}
The reduced-space Koopman operator is linear and has demonstrated that it can accurately reconstruct scenarios such as Kármán vortex street and plumes. However, these datasets feature velocity fields with relatively smooth variations over time. To test whether our method can adapt to scenarios with abrupt changes in the velocity field, we selected a more challenging scenario: colliding vortex rings. In this experiment, two point vortices are initialized and collide head-on. When the vortices meet, the velocity field undergoes a sudden change, resulting in finer vortex structures. We tested our method on this dataset using 150 basis functions. As shown in \reffig{fig:teaser}, our method can reconstruct the transition from the two point vortices before the collision to the rapid formation of a divergent velocity field during the impact, as well as the subsequent emergence of numerous vortical structures around the periphery. This experiment shows that our method can effectively handle datasets with significant and sudden velocity variations.

\subsection{Independence on Simulation Schemes}
Our reduced simulation method is inherently \emph{simulator-agnostic}, allowing it to work seamlessly with a variety of fluid solvers. This flexibility arises from the fact that our method models only the transitions between successive fluid states in the reduced space, rather than being tied to the specific equations or numerical schemes of a given solver. This allows us to apply our method to any fluid simulator without the need for additional adjustments. For instance, in our experiments, we used both the MacCormack~\cite{selle2008unconditionally} + Reflection~\cite{zehnder2018advection} (MC+R) solver and a Lattice Boltzmann Method~\cite{chen1998lattice} (LBM) solver. They are fundamentally different in their discretization and numerical operations. Specifically, the LBM solver does not directly solve the Navier-Stokes equations; instead, it solves the approximated Boltzmann equation, which models fluid dynamics at the mesoscopic scale using particle distribution functions. Our method overcame these challenges by relying solely on the data produced by the solver. In other words, our method is \emph{equation-free}. We detailed the base simulators used for each example in \reftab{tab:experiment_setup}, where our approach successfully reconstructs the velocity fields in all cases. 

In contrast, previous data-driven methods such as \cite{treuille2006model, kim2013subspace} developed a specific numerical scheme for the subspace constructed from data, essentially binding the formulation to one specific base simulation framework. As a result, when switching to a different solver, their method also requires corresponding adjustments. This will introduce significant limitations for users, as they can only input datasets corresponding to a specific solver. In particular, this limits artists usage to only solutions of such solvers, whereas our method accepts hand-modified, or even real-world measured data.

\begin{table*}[ht]
    \centering
    \rowcolors{2}{white}{gray!20} 
    \begin{tabular}{c|c|c|c|c|c}
    \textbf{Examples} & \textbf{Resolution} & \textbf{Dim.} & \textbf{B.C.} & \textbf{Amount of Basis} & \textbf{Base Simulator} \\
    \textbf{Smoke Ring} ($\triangleright$ \reffig{fig:teaser}) & $128 \times 128 \times 256$ & 3D & Open & $150$ & MC+R \\
    \textbf{Rayleigh–Taylor Instability ($\triangleright$  \reffig{fig:dmdadvectioncomparison})} & $1024 \times 512$ & 2D & Dirichlet & $100$ & MC \\
    \textbf{Plume} ($\triangleright$ \reffig{fig:basis_evaluation}) & $128 \times 128 \times 256$ & 3D & Open & $2, 9, 28, 61, 105, 130, 150$ & MC\\
    \textbf{Plume w/ Sphere} ($\triangleright$ \reffig{fig:basis_evaluation}) & $128 \times 128 \times 256$ & 3D & Dirichlet & $2, 9, 28, 61, 105, 130, 150$ & MC \\
    \textbf{Plume w/ Bunny} ($\triangleright$ \reffig{fig:basis_evaluation}) & $128 \times 128 \times 256$ & 3D & Dirichlet & $2, 9, 28, 61, 105, 130, 150$ & MC\\
    \textbf{Reversibility} ($\triangleright$ \reffig{fig:reverse_simulation}) & $512 \times 512$ & 2D & Open & $20$ & MC \\
    \textbf{Editing (K\'arm\'an Vortex Street)} ($\triangleright$ \reffig{fig:karman_editing}) & $512 \times 512$ & 2D & Periodic & $100$ & LBM-BGK \\
    \textbf{Editing (Bunny)} ($\triangleright$ \reffig{fig:bunny_editing}) & $128 \times 128 \times 256$ & 3D & Dirichlet & $50$ & MC \\
    
    \end{tabular}
    \caption{\textbf{Breakdown of Experiment Setup.} The result and experiment setup are detailed in this table, including grid resolution, dimensionality, boundary conditions (B.C.), the number of basis functions used and the base simulator for each result. As for the base simulators, we employ the MacCormack \cite{selle2008unconditionally} (MC), the MacCormack + Reflection \cite{zehnder2018advection} (MC+R) and the Lattice Boltzmann Method with the Bhatnagar-Gross-Krook collision model (LBM-BGK)\cite{chen1998lattice}.}
    \label{tab:experiment_setup}
\end{table*}
\section{Application: Harnessing the Linearity}
\label{sec:application}
Leveraging the \emph{linearity} of DMD operator, as well as the intuition of bases exposed by the spectral decomposition, we have developed several novel applications that extend the capabilities of our Koopman-based reduced-order simulation pipeline. In this section, we explore these applications, demonstrating that our method's unique strengths translate into practical tools for graphics and simulation.

\subsection{Direct Editing Temporal Dynamics}
\label{sec:editing}
\begin{figure}[!ht]
    \centering
    \includegraphics[width=1\columnwidth]{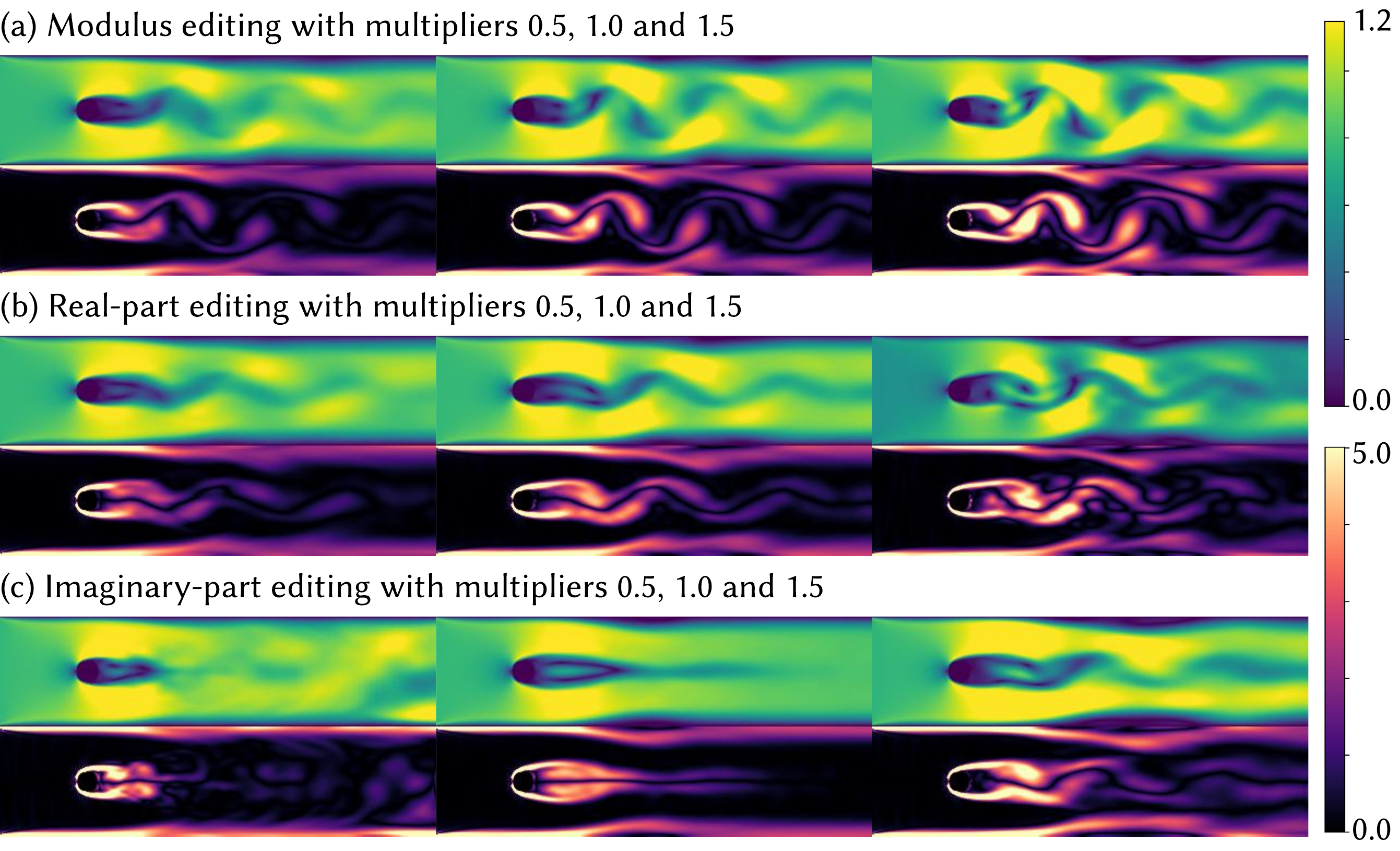}
    \caption{\textbf{Editing temporal dynamics of K\'arm\'an Vortex Street with the Koopman Operator Approximation}. The modifications are applied to the DMD basis coefficients: (a) Scaling the modulus of the DMD basis by factors of 0.5, 1.0, and 1.5, affecting overall amplitude; (b) Adjusting the real part of $\bm{\Omega}$, influencing growth and decay rates of modal contributions; (c) Modifying the imaginary part, altering phase dynamics and wave propagation characteristics. }
    \label{fig:karman_editing}
    \Description{}
\end{figure}

\begin{figure*}[!ht]
    \centering
    \includegraphics[width=1\linewidth]{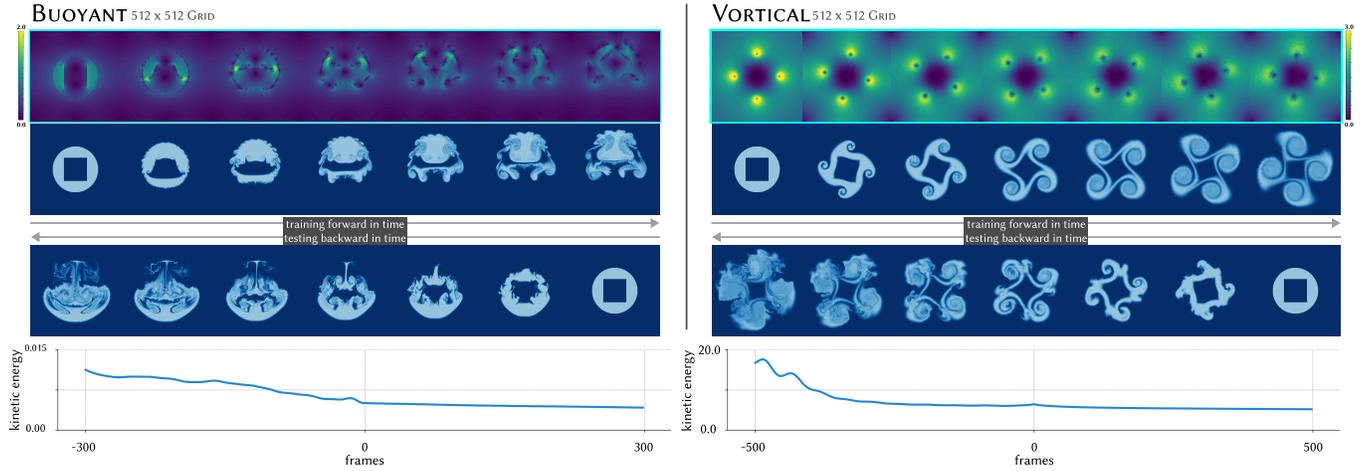}
    \caption{\textbf{Reversibility of Flows with Inversed DMD Operator}. We compare the reconstruction of two distinct fluid flows using Dynamic Mode Decomposition (DMD). The top row in each panel shows the velocity L2-norm of the field used to train the DMD, while the second and third rows depict the temporal evolution of the reconstructed flow fields as applied to an initial density field. The forward-time training phase is followed by a backward-time testing phase to assess predictive accuracy when advecting backward in time. The bottom plots show the evolution of kinetic energy over time. From the buoyant case, we observe the inverted DMD operator $\bm{A^{-1}}$ can still reasonably trace backward in time without compromising much visual quality. The vortical case exhibits a more challenging example where the symmetry should be reconstructed backwards in time. We see that the inverse operator indeed recovers this symmetry, with some acceptable levels of incurred noise. Bottom plots show the evolution of the total kinetic energy over time, demonstrating that our inverse operator actually correctly reverses the arrow of time, reversing the dissipation-related entropy increase over time. Decreasing kinetic energy also validates the \emph{physical plausibility} of our result.}
    \label{fig:reverse_simulation}
    \Description{}
\end{figure*}

Since our method approximates \refeq{eqn:euler_equations} with a linear operator in the full space, this allows us to transform the operator acting on the velocity field into the evolution of different modes under a linear operator. Therefore, we can directly edit the temporal dynamics of the fluid system by modifying the modes of the reduced \koopman{} $\bm{\hat{K}}$:
we set $t_0$ to be the initial time, $\bm{\Omega} = \nicefrac{\log(\bm\Lambda)}{\Delta t}$, where $\Delta t$ is the time step of the dataset. With this, we can rewrite \refeq{eqn:reduced_koopman_simulation} in the following form:
\begin{equation}
    \begin{aligned}
    \bm{u}(t_0 + k\Delta t) &= \bm{\Phi}\exp(\bm{\Omega} t) \bm{z}(t_0) \\
    &= \bm{\Phi}\exp{\left(k(\log(r) + i\theta)\right)} \bm{z}(t_0) \\
    &= \sum_{i = 1}^{n} {w_i} \bm{\Phi_i} r_i^k \left(\cos(k\theta_i) + \sin(k\theta_i)\right) \bm{z_i}(t_0)\\
    \end{aligned}
    \label{eqn:edit_temporal}
\end{equation}
where ${w_i}$ is a user-defined scalar weight, $r_i = \sqrt{\Re(\lambda_i)^2 + \Im(\lambda_i)^2}$ is the \emph{modulus} and $\theta_i = \arctan\left(\Im(\lambda_i), \Re(\lambda_i)\right)$ is the \emph{phase} of the $i$-th eigenvalue $\lambda_i$ in the diagonal \emph{complex} eigenvalue matrix $\bm{\Lambda}$. Notice that this implies that the modes of the spectral decomposition represent different scales of vorticity, completing the physical intuition of the reduced space modes.

As shown in \refeq{eqn:edit_temporal}, our method decomposes a simulation sequence into modes with different growth/decay rates and frequencies.
The growth/decay rate of a mode is reflected in $r_i$, where a larger $r_i$ indicates a higher growth rate (or a lower decay rate), and vice versa.
The frequency of a mode is represented by the absolute value of $\theta_i$, with a larger absolute value corresponding to a higher frequency mode, and vice versa.
Furthermore, the different modes are decoupled, allowing for the adjustment of the relative proportions between modes.
As a result, these properties provide the artist with powerful tools to edit the simulation playback. The artist can modify the overall velocity field by adjusting the proportion ($w_i$), growth/decay rate ($r_i$), and frequency ($\theta_i$) of specific modes.
In the experiments, we directly adjust the real part of $\bm{\Omega_i}$ to control $r_i$, modify the imaginary part of $\bm{\Omega_i}$ to control $\theta_i$, and vary the modulus of $\bm{\Phi_i}$ to control $w_i$.
\paragraph{Editing the K\'arm\'an Vortex Street}
The first example is editing on the classic K\'arm\'an vortex street. We filter the imaginary part of $\bm{\Omega}$ and cluster modes with an absolute value smaller than $0.01$ as \emph{low-frequency cluster}, and the rest as \emph{high-frequency cluster}.
The low-frequency mode manifests as a laminar flow, with its phase changing very slowly over time. The high-frequency mode is represented by vortical structures distributed on both sides of the cylinder, where the phase of this mode changes relatively quickly over time.
As seen in \reffig{fig:karman_editing}, when we adjust the modulus of the high-frequency cluster from $0.5$ to $1.5$, the intensity of the vortices increases, which is as we expected. When we set the real part of $\bm{\Omega}$ to $0.5$, it can be observed that the high-frequency motion decays faster than user input. When we set the real part of $\bm{\Omega}$ to $1.5$, it can be observed that the high-frequency motion decays slower than user input. Similarly, when we tune the imaginary part of $\bm{\Omega}$ from $0.5$ to $1.5$, we could observe the oscillation frequency of the fluid trail transitions from slow to fast compared to user input.
\paragraph{Editing the Plume with Bunny}
To evaluate the editing capability of our method, we scale our editing scenario to 3D. With the same filtering procedure as in the K\'arm\'an vortex street example, we set the low-frequency cluster to high-frequency cluster ratio to $4:1$, $2:1$, $1:2$, and $1:4$, and compared the results with the user input. From the results, we observe that when the proportion of low-frequency cluster is increased, with a ratio of $4:1$, the top of the plume lacks "wrinkles" and appears more "fluffy". This is because the velocity field is dominated by smoother, lower-frequency modes than the original user input. Conversely, when the proportion of high-frequency cluster is increased, with ratios of $1:4$, the plume developes more detailed plume structure around the top, as the velocity field now emphasizes more high-frequency details compared to the user input.

\subsection{Reversibility of the Reduced Simulation}
Although physically-based fluid simulations have the capability to generate stunning visuals, when artists aim to direct the fluid's evolution toward a predefined target shape, challenges arise. It is a long standing problem in the community that people aim to enable users with \emph{spatial control}. In this example, we aim to enable users to do \emph{temporal control}, motivated by a prior work \citet{oborn2018time}. Compared to previous work \shortcite{oborn2018time} where the authors employ a self-attraction force to replace the arbitrary external forces, providing a stable, physics-motivated, but time-consuming approach, we propose a data-driven, fast, and easy to implement method to address the same problem.

\label{sec:reversibility}

We observe that that given $\bm{\tilde{K}} = \bm{\Phi} \bm{\Lambda} \bm{\Phi}^+$, we could easily compute the \emph{inverse} of the truncated \koopman{} $\bm{\tilde{K}}^{-1} = (\bm{\Phi} \bm{\Lambda} \bm{\Phi}^+)^{-1} = \bm{\Phi} \bm{\Lambda}^{-1} \bm{\Phi}^+$, which is essentially the approximate inverse time evolution $\bm{f}^{-1}(\bm u)$ of the fluid system. This allows us to reverse the simulation by applying the inverse truncated \koopman{} to the current state of the fluid system:
\begin{equation}
    \label{eqn:reverse_simulation}
    \begin{aligned}
        \bm{u}(t) &= \bm{A}^{-1} \bm{u}(t + \Delta t), \\
        \bm{u}(t) &= \bm{\Phi} \bm{\Lambda}^{-1}\bm{\Phi}^+ \bm{u}(t + \Delta t), \\
        \bm{u}(t) &= \bm{\Phi} \bm{\Lambda}^{-1} \bm{z}(t + \Delta t).
    \end{aligned}
\end{equation}

Similar to \refeq{eqn:reduced_koopman_projection}, we could train the reduced \koopman{} on the forward simulation data and then apply the inverse reduced \koopman{} to reverse the simulation, given a state of the fluid system.

\begin{figure*}[!ht]
    \centering
    \includegraphics[width=1\linewidth]{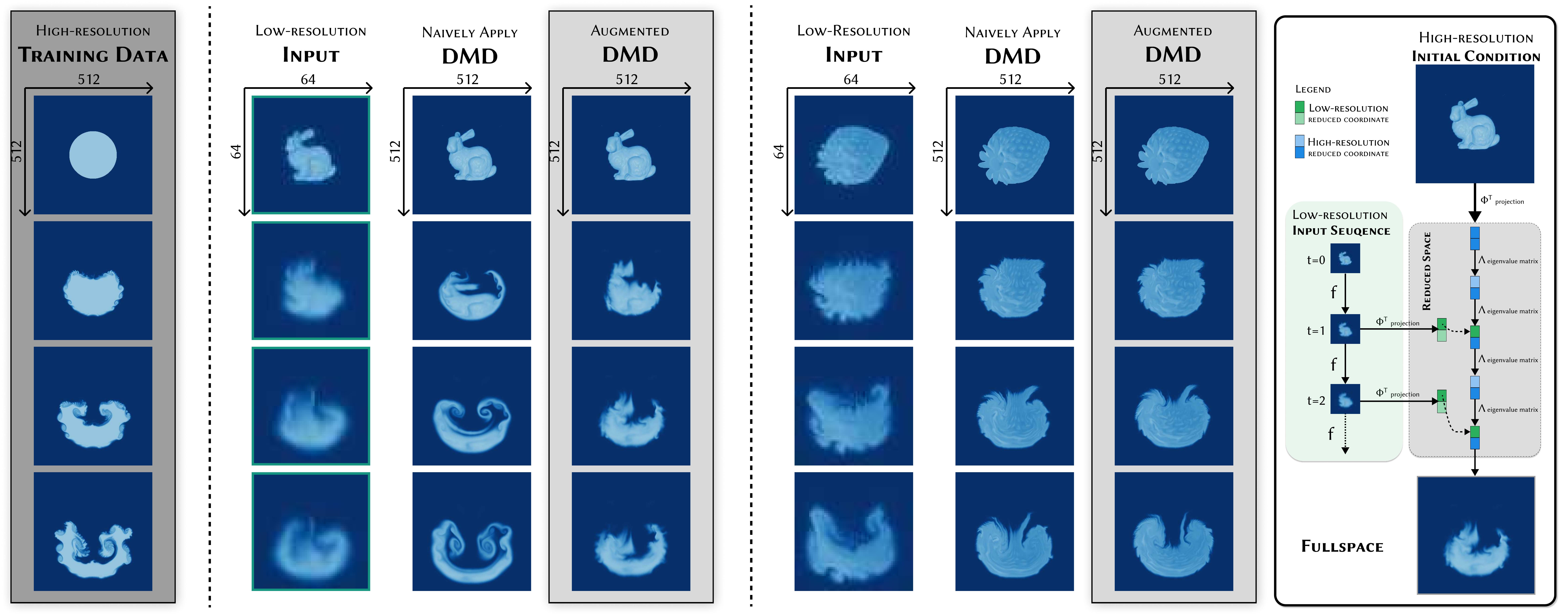}
    \caption{\textbf{Upsampling and Generalization to Unseen Sequences with Trained DMD Operator}. Two different input low-resolution fluid simulations (bunny and SIGGRAPH logo) are upscaled using the same DMD operator trained on a different velocity field. Initial velocity fields are seeded as moving down based on the input density field.    
    Naive application of DMD shown in each middle column, and our \emph{augmented DMD upresolution} method shown on the right columns. 
    Schematic of our method presented on the far right. At each frame, we project the low-resolution artist-directed input into the low-order bases of our reduced representation, using these to replace the low-order terms of the DMD field. Notice that naive application of DMD simply moves towards the known input training data, while our augmented field matches the low-resolution input more closely, with extra high-order detail gained from the DMD operator.}
    \label{fig:upsample}
    \Description{}
\end{figure*}

\paragraph{Reversibility of Buoyant Flow}
We experiment our approach on a simple buoyant flow setup (\reffig{fig:reverse_simulation}, left). Our dataset was initialized with a density field shaped like the SIGGRAPH logo, with the density value set to 1. A density value of $1$ density field was driven by a velocity field where an upwards velocity of $0.3$ is set within the SIGGRAPH logo and downwards elsewhere. We run the simulation for $300$ frames to construct the dataset, and trained the DMD operator on this dataset. The inverse operator $\bm{\tilde{K}}^{-1}$ was then applied to the initial velocity field of the dataset at $t=0$ (frame $0$). By iteratively applying the inverse operator, we obtained the velocity fields for the preceding frames, starting from frame $-1$, frame $-2$, and all the way back to frame $-300$.
When examining the evolution of the density field from frame -300 to frame 300, it is evident that the velocity field remains consistently upward and smooth, indicating that our method is both reasonable and effective.
Further analysis of the energy of the velocity field obtained through the inverse process and the velocity field from the dataset reveals a downward trend in energy, with a smooth and reasonable curve, consistent with fluids with dissipative properties. This demonstrates that our inverse operator has the ability to predict a \emph{physically-plausible} velocity field prior to the dataset.

\paragraph{Reversibility of Vortical Flow}
To challenge the method with a scene of nontrivial vortical structure, we initialized a vortex sheet by placing four vortices at the corners of the domain (\reffig{fig:reverse_simulation}, right). We generated the dataset using the same procedure as in the previous experiment, resulting in a collection of $500$ frames. Subsequently, we constructed the inverse operator to recover the velocity fields preceding the dataset.
The results show that the density field (counterclockwise) and the dataset (clockwise) rotate in the opposite direction, which indicates that the velocity field predicted by the inverse operator is correct. This is because the vortex sheet velocity field continuously rotates in a clockwise direction, and by examining the density field from frame -500 to frame 500, we observe that the field indeed undergoes continuous clockwise rotation.
From the energy field analysis, the results show that, except for the significant energy fluctuation between frames -500 and -450, the energy consistently decreases in the remaining frames, with a consistent slope. This further demonstrates the robustness of our method.

\subsection{Upsampling with Reduced Koopman Operator}

The scale of the imaginary part of eigenvalues in $\bm{\Lambda}$ encode different scales of turbulent modes, enabling us to use a trained DMD operator to add in secondary motion to an existing fluid simulation. This is particularly useful for \emph{upscaling} a low-resolution fluid, simulated using stable fluid for example, leveraging the DMD basis to add in turbulent modes that were too small for the low-res sim to capture. This upscaling problem has been explored in prior work \cite{kim2008wavelet, nielsen2009guiding}, but we show that due to the linearity of the Koopman operator, and the physical intuition on each of its reduced bases, this upscaling is essentially attained for \emph{free}, amounting to nothing more than a linear combination of two matrix multiplications. 

\subsubsection{Evolution} \label{sec:upres_direct}

Suppose we have frames of a low-res input velocity field $\{\bm{L}_0, \bm{L}_1, \bm{L}_2, \dots, \bm{L}_T\}$, a high-res initial condition $H_0$. Additionally, we have some DMD basis $\bm{\Phi}$ trained on some high-res simulation distinct from the low-res simulation, with corresponding eigenvalues $\bm{\Lambda}$, sorted by the length of their imaginary parts in increasing order. At the first frame, we can generate the reduced-space initial condition by simply using our basis mapping $R_0 = \bm{\Phi}^TH_0$.

Now, for every subsequent frame $t$, we generate $R_t$ by first applying the DMD evolution on the previous reduced space frame to produce an intermediate state $R^*_t=\bm{\Lambda}R_{t-1}$. We also produce a representation of the current frame of the low-res input in reduced space $P_t = \bm{\Phi}^TL_t$. Now, we have a representation of the \emph{current} frame of the low-res input, and the DMD \emph{time evolution} of the \emph{previous} reduced space frame. We want to keep the low-order bulk flow of the low-res input, and augment it with the high-order turbulent flow learned by the DMD basis. To that end, we split each reduced-space vector into a low-order and high-order part: $R^*_t = \left[R_t^{*L}\ R_t^{*H}\right]$, $P_t=\left[P_t^L\ P_t^H\right]$. Now, we take only the low-order modes of the input flow, and the high-order modes of the DMD-evolved flow, to produce our new reduced space velocity field $R_t=\left[P_t^L\ R_t^{*H}\right]$. From here, we can just apply the basis to return to high-resolution full-space $H_t=\bm{\Phi}R_t$.

We note that the composition operators here are linear. We can simply represent them with selection matrices $S^H$, $S_L$, for the high- and low-order bases respectively, such that $R_t=S^LP_t + S^HR_t^*$. Since the DMD operator is also linear, we note that this entire upscaling method is linear by construction.

Results are shown on \reffig{fig:upsample}. We see that even if the initial velocity field is significantly different from the input field, the low-order basis is able to capture the bulk flow of the low-resolution input, and modify the DMD-produced field accordingly. In particular, we note that naively applying the DMD operator, without passing the low-resolution input field into the low-order bases, ends up reconstructing the original training set, rather than a velocity field directed by our input. This is demonstrated by the results for the two initial conditions being very similar, whereas our augmented field matches the input much closer.

\subsubsection{Projection}

The above governs the time evolution of the velocity field. In some cases, where the input velocity field differs significantly from the training data used for the DMD basis, the above as written will still produce velocity fields that are unacceptably different from the input velocity field. This is largely representation error, fields that are far away from the training data are less representable by the reduced space. In these cases, we can again leverage our input low-res field, this time as a constraint. 

Essentially, we would like to project our velocity field $\bm{H}_t$ onto the space of velocity fields that are identical to the input low-res field when downsampled to that resolution. This can be represented as an equality-constrained quadratic problem,
\begin{align}
    &\argmin_x \frac{1}{2}(\bm{x}-\bm{H}_t)^T(\bm{x}-\bm{H}_t) \\
    &\text{subject to } \bm{Ax} = \bm{L}_t,
\end{align}
where $\bm{A}$ is a downsampling operator that converts from high-res to low-res. 
Notice that because the downsampling operator does not change for the duration of the simulation. Thus, the KKT (Karush-Kuhn-Tucker) matrix can be precomputed making the projection a single matrix multiply during runtime.

\begin{wrapfigure}{r}{0.5\columnwidth}
    \vspace{-12pt}
    \includegraphics[width=0.5\columnwidth]{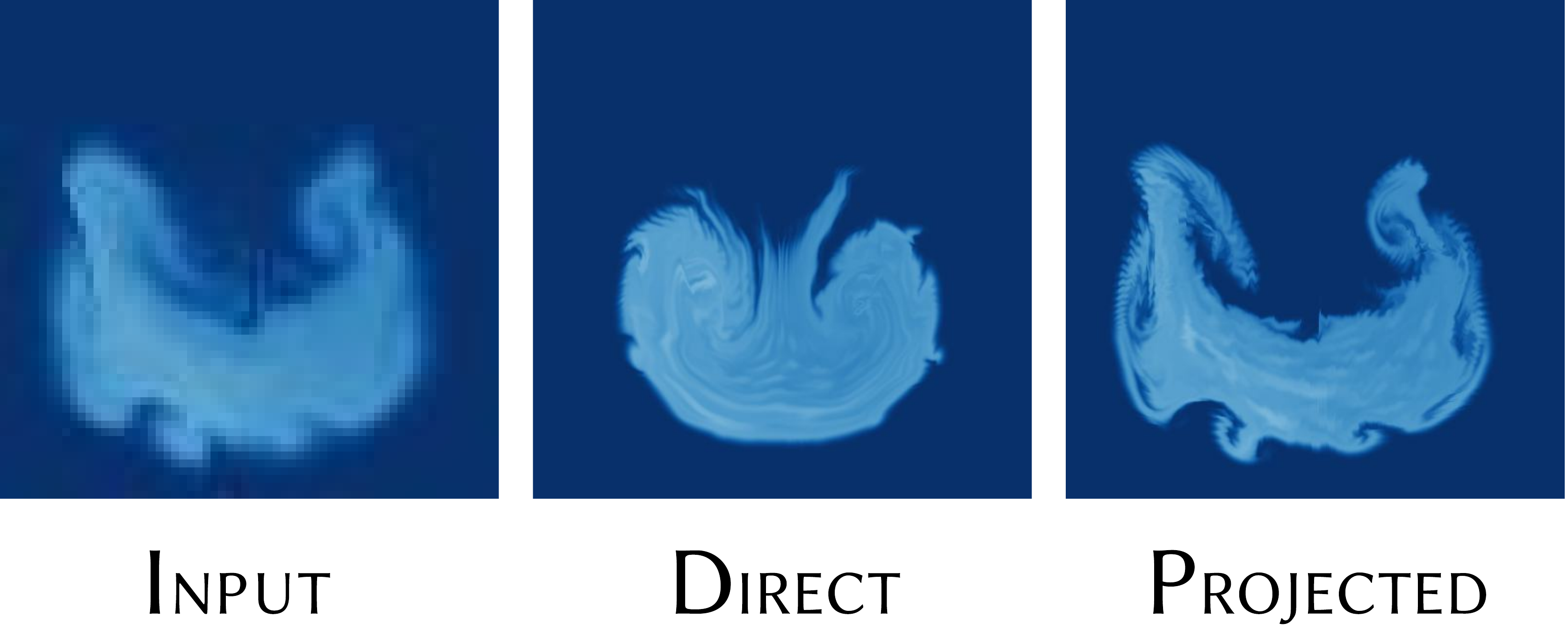}
    \hspace{-2pt}
    \vspace{-20pt}
    \label{fig:qp_project}
\end{wrapfigure}

As a sanity check, we show the effect of this projection here: it is apparent with the projection step,we can recover fields that are much closer to the input, yet retaining extra high-order detail. And, of course, because these are all linear, linear combinations of the direct and projected fields can be taken. In particular, because the basis functions of reduced space are orthogonal, a diagonal matrix of linear weights can be taken, preferring projected for low-order modes and direct for high-order modes, for example.
\section{discussion}
\label{sec:discussion}
DMD, similar to PCA, performs dimensionality reduction on some dataset, and provides a reduced basis for representing the fluid flow. The key difference is that in constructing this basis, DMD optimizes for the \emph{action} of some linear operator that maps the data from one time state to the next, while PCA only looks at the state snapshots individually. This temporal information allows DMD to find a reduced space that can represent the evolution of the field as a linear operator within this reduced space, as opposed to PCA that only seeks to find a reduced space that best minimizes reconstruction error.

We also note that as a spectral decomposition of the reduced Koopman operator, much of the \emph{directability} power of DMD comes from the exposure of the complex-valued eigenvalues. Because each eigenvalue represents a different speed of rotation, and their corresponding eigenvectors represent their spatial bases, we gain manipulability over different scales of turbulence for free. This is similar in concept to other spectral methods \cite{kim2008wavelet}, but with corresponding spatial bases that learned from input data rather than predefined bases.
We thus show that DMD straddles the realm between \emph{spatial} order reduction (PCA and co.) and \emph{temporal} order reduction (spectral methods), gaining strengths of both methodologies.

The other interesting observation is that \textit{dynamic mode decomposition} (DMD) shares the same algebraic backbone as a first-order \textit{vector autoregression}, VAR(1)~\cite{stock2001vector}: both estimate a linear map that advances a state from $t$ to $t{+}1$. Under a \textit{noise-free} assumption, the two time-shifted snapshot matrices $X'$ and $X$ introduced in \refeq{eqn:reduced_koopman} satisfy the VAR(1) relation
\begin{equation}
    \label{eqn:var}
    X' = \bm A X,\quad \bm A\in\mathbb R^{(Nn)\times (Nn)},
\end{equation}
with $\bm A$ interpreted as the \emph{state-transition matrix}.  DMD solves the identical least-squares problem but replaces the full $\bm{A}$—the restricted Koopman operator $\bm{K}$ in our notation—with its truncated SVD, yielding a compact low-rank approximation that captures the dominant dynamical modes.

\subsection{Limitations}
\label{sec:limitation}
\paragraph{SVD for Dimensionality Reduction}
Like other low-rank approximation approaches, DMD requires performing SVD for dimensionality reduction. This demands high memory consumption for large-dimensional data, such as 3D fluid simulation, on the order of $\mathcal{O}(n^2)$, where $n$ is the number of state variables. Despite applying randomized SVD to reduce this memory footprint, larger simulations are still costly to train. In the future, we will explore the use of other dimensionality reduction methods, such as autoencoders, to reduce the dimensionality of the dataset.

\paragraph{Unseen External Force Generalization} It is well-known that reduced-order model trades \emph{generalization} for \emph{performance}; our method is no exception. We have demonstrated that we can represent velocity fields that were unseen by DMD, as well as interactively push current states towards modified states. Despite this, representability is still a key limitation--there just may not be the spatial basis and the reduced \koopman{} to support certain velocity fields and their nonlinear evolution. Thus, users may find that interacting with the fields, especially in states far away from observed data, behave strangely. This warrants further work on improving spatial generalization of these methods.

\subsection{Future Work}
\label{sec:future_works}
We have identified various potential avenues for extension, particularly ones that exploit the linearity of the DMD operator.

\paragraph{Augmenting Fluid State with Density Field}
The DMD framework allows us to trivially evaluate the velocity field at any point in time via a low-dimensional matrix exponentiation (as shown in \reffig{fig:dmdadvectioncomparison}).
unfortunately evaluating an immersed  \textit{density field } at any point in 
time is not so trivial.
To achieve the time-evolved density field, we've first had to resort to classical numerical integration, first evaluating the velocity field with our fast DMD model, and then using that velocity field to advect the density field forward.
As an exciting avenue of future work, this could potentially be sidestepped by defining the fluid state as including the density field (as opposed to just the velocity field), and forming a \koopman{} on this augmented system. This would allow us to use the exact same DMD framework to quickly query the density of the flow at any point in time.

\paragraph{Adaptive Blending between DMD Operators}
As a linear operator, linear combinations of DMD operators construct a valid linear operator. Therefore, we can train multiple DMD operators and flexibly combine them to obtain a richer set of new motion patterns.
Similar to building an ensemble, once we have multiple pre-trained DMD operators, we solve a minimal optimization problem to adaptively selects the optimal linear combination of DMD operators based on user input, achieving the desired motion effects in the simulation.
\paragraph{Inverse Design}
Since the DMD operator decomposes the energy information ($\bm{\Phi}$) and frequency information ($\bm{\Lambda}$) of a simulation sequence, we can adjust the eigenbasis of the DMD operator based on the user-defined objectives, thereby modifying the energy and frequency information represented by the DMD operator. We take gradients of the DMD operator with respect to some target images at prescribed points in time, and use this as an optimizer to construct a \emph{new} DMD operator that represents flow that evolving through the specified snapshots.
\paragraph{Learning Control Space Enrichment}
Inspired by DMD with control \cite{proctor2016dynamic}, we could train a neural network to enrich the linear control space of the DMD operator with nonlinear responses. This network could be a direct way to expand the expressiveness of the DMD operator, allowing us to learn more diverse motion patterns and handle more complex user inputs.
\section{Conclusion}
We introduce \emph{Dynamic Mode Decomposition} (DMD) to the field of graphics. Unlike previous methods that approximate the solution space, our approach is an \emph{equation-free}, \emph{data-driven} method that learns a low-rank approximation of the state transition matrix directly from the dataset. Its linear nature enables rapid reconstruction of the dataset and shows superior reconstruction quality compared to previous data-driven methods. 
The operator linearity also unlocks a collection of artist-centric graphics applications: (1) by decomposing the dataset into a linear combination of temporally oscillating modes of different frequencies, our method enables artists to edit the dataset by adjusting the mode's modulus, growth/decay rate, and frequency, all without requiring knowledge of the solver's details; (2) with the inverse of the \koopman{}, we can go back in time and rollout animations at negative frame ranges; (3) users can supplement low-resolution flow sequences and generate new enriched flow details with the trained DMD operator, taking advantage of the high-frequency information stored within the operator.

\begin{acks}
    We are grateful to Jonathan Chalaturnyk for illuminating discussions that shaped the early stages of this research. We thank Xinwen Ding for helpful proofreading feedback. This research was made possible with the administrative support of our lab’s system administrator, John Hancock, and financial officer, Xuan Dam. This research was supported by the Natural Sciences and Engineering Research Council of Canada (NSERC) through grant RGPIN-2021-03733, whose funding made this work possible.
\end{acks}

\bibliographystyle{ACM-Reference-Format}
\bibliography{references}

\appendix
\section{Koopman Theory} \label{sec:koopman_theory}

We briefly summarize Koopman theory, following the presentation in \citet{modern_koopman_theory}, to motivate the the linearization introduced in \refeq{eq:linearization}. Once again, consider the autonomous system:
\begin{align}
    \frac{d\bm{u}}{dt}=\bm{f}(\bm{u}). \label{eqn:autonomous}
\end{align}
The time-dependent state $\bm{u}(t)$ subject to this dynamical system and some initial condition $\bm{u}(0)=\bm{u}_0$ may be represented with a time-dependent family of flow maps $\{\bm{F}_t\}_{t>0}:\mathbb{R}^{Nn} \to \mathbb{R}^{Nn}$ such that:
\begin{align}
    \bm{u}(t)=\bm{F}_t(\bm{u}_0).
\end{align}
That is, it maps some initial state $\bm{u}_0$ from time $0$ to its state at time $t$.
In general, $\bm{F}(\bm{u})$ is not a linear map, but as per \citet{koopman1932dynamical} may be lifted in a higher space that admits a linear mapping in that space. In particular, take a Hilbert space $\mathcal{H}\subseteq\{g:\mathbb{R}^N\to \mathbb{C}\}$ of scalar observation functions $g\in\mathcal{H}$. With these observables, \citet{koopman1931hamiltonian} defines a family of \emph{Koopman operators} $\bm{\mathcal{K}}_t: \mathcal{H}\to\mathcal{H}$ such that

\begin{align}
\label{eq:Ktdef}
    \bm{\mathcal{K}}_tg = g \circ \bm{F}_t, \forall g \in \bm{\mathcal{H}},
\end{align}
forming a trajectory in $\mathcal{H}$.
That is to say, $\bm{\mathcal{K}}_t$ maps the measurement operator $g$ to its state at time $t$. Consequently, rather than following the nonlinear flow according to $\bm{F}_t$ of some initial state $\bm{u}_0$, we can instead take an observation of $\bm{u}_0$, and linearly evolve it according to $\mathcal{K}_t$. Notice that because $\mathcal{H}$ is a linear space, it follows from \refeq{eq:Ktdef} that:
\begin{align}
\label{eqn:linear}
    \mathcal{K}_t(\alpha g_\alpha + \beta g_\beta) = \alpha \mathcal{K}_t g_\alpha + \beta \mathcal{K}_t g_\beta,
\end{align}
\emph{irrespective} of whether $\bm{F}_t$ is linear in state space $\mathbb{R}^N$.

An infinitesimal generator $\mathcal{L}$ of the family $\{\mathcal{K}_t\}$ exists given $f$ is an autonomous system (\refeq{eqn:autonomous}), and can be found via the limit,

\begin{align}
\label{eqn:limit}
    \mathcal{L}(g(\bm{u}(t))) = \lim_{\tau\to 0}\frac{\mathcal{K}_\tau g(\bm{u}(t))-g(\bm{u}(t))}{\tau}.
\end{align}

Notice that the limit is exactly the definition of a derivative, so that:
\begin{align}
\label{eqn:deriv}
    \frac{d}{dt}g(\bm{u}(t))=\mathcal{L}g(\bm{u}(t)).
\end{align}
This is analogous to \refeq{eqn:autonomous}, but with the finite nonlinear operator $\bm{f}$ replaced by an infinite-dimensional linear operator $\mathcal{L}$. This has the matrix exponential solution $\bm{g}(\bm{u}(t+\tau)) = e^{\mathcal{\mathcal{L}}(t+\tau)}\bm{g}(\bm{u}(t))$. Thus, for discrete time $t\in\mathbb{N}$, we can simply take some constant timestep $\tau$ to find a single Koopman operator $\mathcal{K}=e^{\mathcal{L}\tau}$, notably acting linearly in $\mathcal{H}$, whose $m$-times repeated application $\{\mathcal{K}^m=\mathcal{K}\circ\mathcal{K}\circ\cdots\circ\mathcal{K}\}$ constructs the family of operators at discrete multiples of $\tau$. This discrete-time Koopman operator, often called \emph{the} Koopman operator, thus giving the timestepping scheme:
\begin{align}
    g(\bm{u}((k+1)\tau)) = \mathcal{K}g(\bm{u}(k\tau)).\label{eq:koopman_discrete_time}
\end{align}

Now, suppose we have some input data consisting of a (necessarily) finite set of measurements. We can consider each one of these measurements to be exactly one evaluation of the observation function. We can thus restrict our set of observation functions only to those required for the given measurements. That is, we only need to consider observation functions spanned by the set $\{g_i(\bm{u})=\bm{u}(\bm{x}_i)\}$, where $\bm{x}_i$'s are all the spatial degrees of freedom of our measurements. In other words, we limit $\mathcal{H}$ only to the subspace of observables spanned by input data. As a convenient representation of the entire system observation function, we define the stacked vector,
\begin{align}
    \bm{g}(\bm{u}) = \begin{bmatrix}
        g_1(\bm{u}) \\ g_2(\bm{u}) \\ \vdots \\ g_n(\bm{u})
    \end{bmatrix} = \begin{bmatrix}
        \bm{u}(\bm{x}_1) \\ \bm{u}(\bm{x}_2) \\ \vdots \\ \bm{u}(\bm{x}_n)
    \end{bmatrix}
\end{align}
where $n$ is the total spatial degrees of freedom of our measurements.
We then define $\bm{K}$ to be the Koopman operator restricted to this subspace, providing a \emph{linearized} timestepping operator,
\begin{align}
    \bm{g}_{k+1} = \bm{K}\bm{g}_{k},
\end{align}
analogous to \refeq{eq:koopman_discrete_time}.

We have thus shown that the finite-dimensional \emph{nonlinear} system provided in \refeq{eqn:autonomous} may be lifted into a \emph{linear} system in an \emph{infinite-dimensioned} Hilbert space $\mathcal{H}$, then restricted to a finite linear system respecting a finite set of measurements. We would like to emphasize here, as a key limitation of the theory, the requirement of the nonlinear system to be \emph{autonomous}. That is, the dynamics is necessarily only state-dependent and time-independent. This means that we cannot hope to train an operator that includes some time-dependent source term. Consequently, our moduli-editing schemes cannot be seen as solving the original system given by \refeq{eqn:autonomous}, but rather provides a perturbed version of the original solution.
\section{Additional Results}
\begin{figure*}[!htp]
    \centering
    \includegraphics[width=\linewidth]{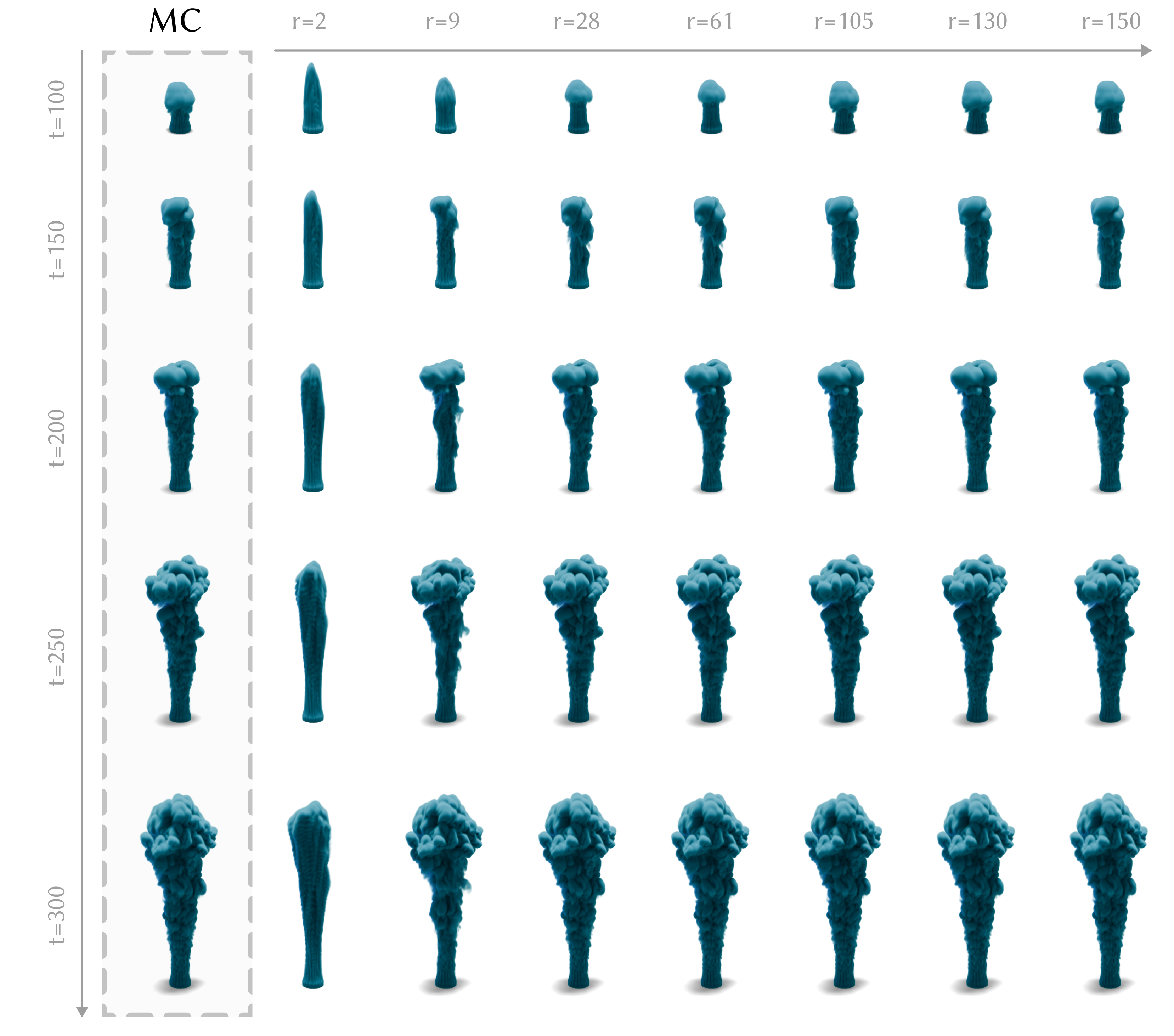}
    \caption{Additional plume simulation results in \reffig{fig:basis_evaluation}.}
    \label{fig:appendix_plume}
    \Description{
        Additional plume simulation results in \reffig{fig:basis_evaluation}.
        Comparison of fluid simulation results using Dynamic Mode Decomposition (DMD) with varying numbers of basis functions ($r$).**  
        The leftmost column (MC) represents the Monte Carlo (ground truth) simulation. The remaining columns depict reconstructions using DMD with increasing numbers of basis functions ($r = 2, 9, 28, 61, 105, 130, 150$). Rows correspond to different time steps ($t = 100, 150, 200, 250, 300$). As $r$ increases, the reconstruction quality improves, capturing more details of the plume-like fluid structure over time.}
\end{figure*}

\clearpage  

\begin{figure*}[t]
    \centering
    \includegraphics[width=\linewidth]{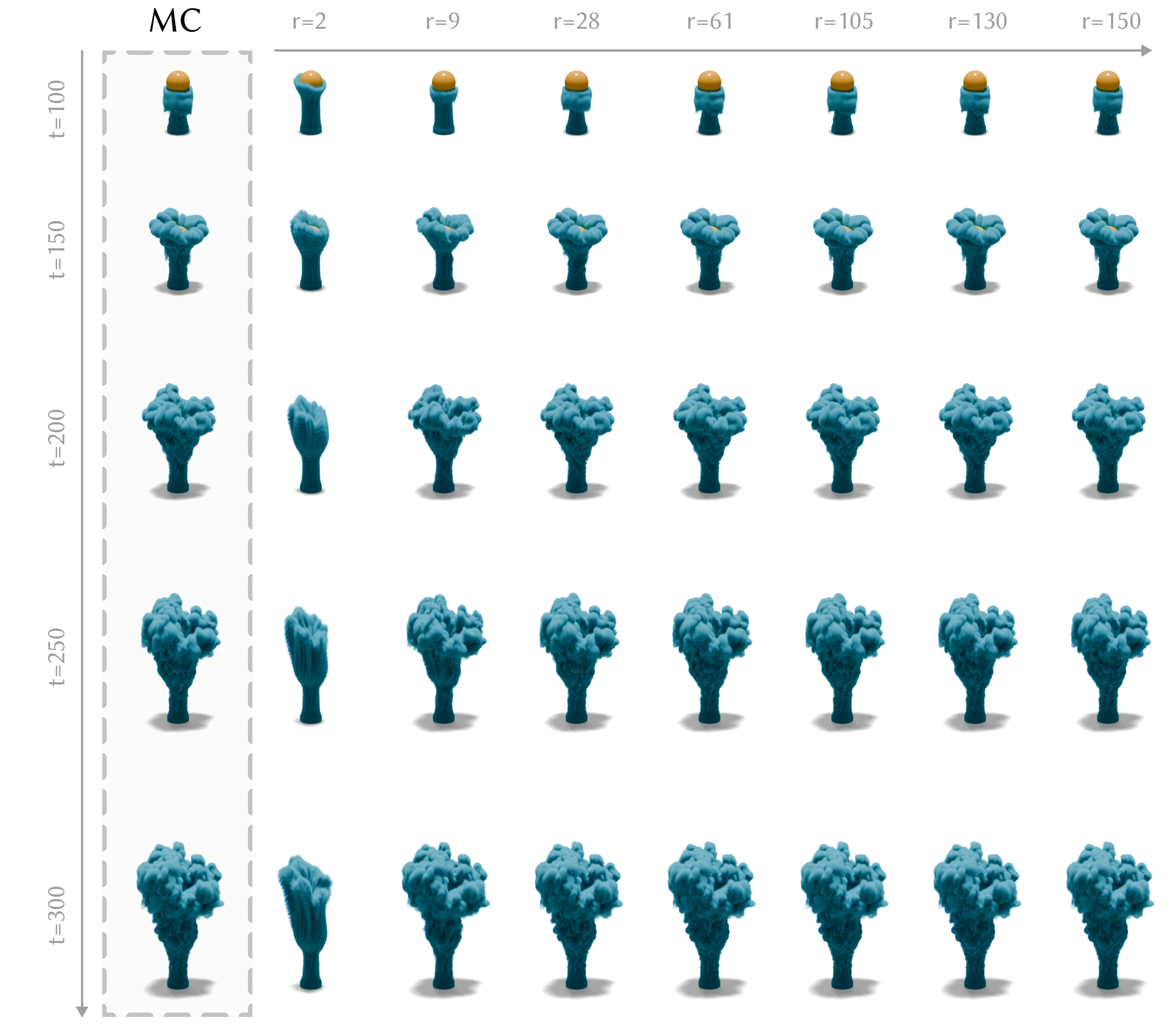}
    \caption{Additional plume with sphere simulation results in \reffig{fig:basis_evaluation}.}
    \label{fig:appendix_sphere}
    \Description{
        Additional plume with sphere simulation results in \reffig{fig:basis_evaluation}.
        Comparison of fluid simulation results using Dynamic Mode Decomposition (DMD) with varying numbers of basis functions ($r$).**  
        The leftmost column (MC) represents the Monte Carlo (ground truth) simulation. The remaining columns depict reconstructions using DMD with increasing numbers of basis functions ($r = 2, 9, 28, 61, 105, 130, 150$). Rows correspond to different time steps ($t = 100, 150, 200, 250, 300$). As $r$ increases, the reconstruction quality improves, capturing more details of the plume-like fluid structure over time.}
\end{figure*}

\clearpage  

\begin{figure*}[t]
    \centering
    \includegraphics[width=\linewidth]{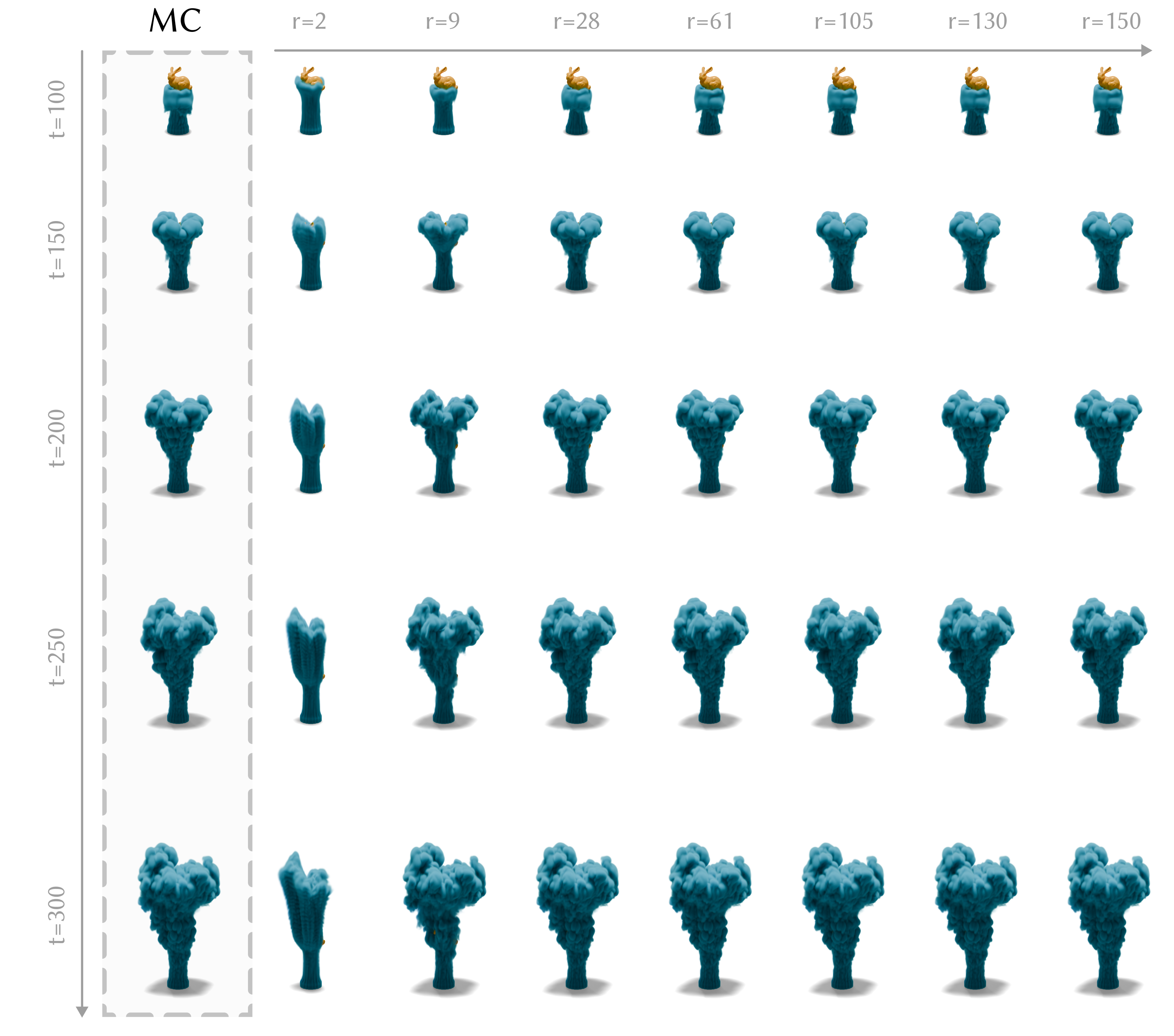}
    \caption{Additional plume with bunny simulation results in \reffig{fig:basis_evaluation}.}
    \label{fig:appendix_bunny}
    \Description{
        Additional plume with bunny simulation results in \reffig{fig:basis_evaluation}.
        Comparison of fluid simulation results using Dynamic Mode Decomposition (DMD) with varying numbers of basis functions ($r$).**  
        The leftmost column (MC) represents the Monte Carlo (ground truth) simulation. The remaining columns depict reconstructions using DMD with increasing numbers of basis functions ($r = 2, 9, 28, 61, 105, 130, 150$). Rows correspond to different time steps ($t = 100, 150, 200, 250, 300$). As $r$ increases, the reconstruction quality improves, capturing more details of the plume-like fluid structure over time.}
\end{figure*}

\clearpage  

\end{document}